\UseRawInputEncoding
\documentclass[10pt,letterpaper]{article}
\usepackage[top=0.85in,left=2.75in,footskip=0.75in]{geometry}

\usepackage{amsmath,amssymb}

\usepackage{changepage}

\usepackage{textcomp,marvosym}

\usepackage{cite}

\usepackage{nameref,hyperref}

\usepackage[right]{lineno}

\usepackage[nopatch=eqnum]{microtype}
\DisableLigatures[f]{encoding = *, family = * }

\usepackage[table]{xcolor}

\usepackage{array}

\newcolumntype{+}{!{\vrule width 2pt}}

\newlength\savedwidth

\newcommand\numberthis{\addtocounter{equation}{1}\tag{\theequation}}

\raggedright
\usepackage[aboveskip=1pt,labelfont=bf,labelsep=period,justification=raggedright,singlelinecheck=off]{caption}
\renewcommand{\figurename}{Fig}

\makeatletter
\renewcommand{\@biblabel}[1]{\quad#1.}
\makeatother

\usepackage{lastpage,fancyhdr,graphicx}
\usepackage{epstopdf}
\usepackage{booktabs}       
\usepackage{subcaption}
\usepackage{multirow}
\usepackage{algorithm}
\usepackage{algpseudocode}
\usepackage[normalem]{ulem}
\fancyheadoffset[L]{2.25in}
\fancyfootoffset[L]{2.25in}
\useunder{\uline}{\ul}{}

\begin{document}
\vspace*{0.2in}

\begin{flushleft}
{\Large
\textbf\newline{Adapting Physics-Informed Neural Networks to Improve ODE Optimization in Mosquito Population Dynamics} 
}
\newline
\\
Dinh Viet Cuong\textsuperscript{1*},
Branislava Lali\'{c}\textsuperscript{2},
Mina Petri\'{c}\textsuperscript{3},
Nguyen Thanh Binh\textsuperscript{4},
Mark Roantree\textsuperscript{5}
\\
\bigskip
\textsuperscript{1} School of Computing, Dublin City University, Dublin, Ireland
\\
\textsuperscript{2} Faculty of Agriculture, University of Novi Sad, Serbia
\\
\textsuperscript{3} Avia-GIS NV, Zoersel, Belgium
\\
\textsuperscript{4} University of Science, Ho Chi Minh City, Vietnam
\\
\textsuperscript{5} Insight Centre for Data Analytics, Dublin City University, Dublin, Ireland
\\
\bigskip

%
%





* Corresponding author: \\
Email: dinhviet.cuong@dcu.ie

\end{flushleft}
\section*{Abstract}
Physics informed neural networks have been gaining popularity due to their unique ability to incorporate physics laws into data-driven models, ensuring that the predictions are not only consistent with empirical data but also align with domain-specific knowledge in the form of physics equations. The integration of physics principles enables the method to require less data while maintaining the robustness of deep learning in modelling complex dynamical systems. However, current PINN frameworks are not sufficiently mature for real-world ODE systems, especially those with extreme multi-scale behavior such as mosquito population dynamical modelling. In this research, we propose a PINN framework with several improvements for forward and inverse problems for ODE systems with a case study application in modelling the dynamics of mosquito populations. The framework tackles the gradient imbalance and stiff problems posed by mosquito ordinary differential equations. The method offers a simple but effective way to resolve the time causality issue in PINNs by gradually expanding the training time domain until it covers entire domain of interest. As part of a robust evaluation, we conduct experiments using simulated data to evaluate the effectiveness of the approach. Preliminary results indicate that physics-informed machine learning holds significant potential for advancing the study of ecological systems.



\section*{Introduction}
\label{sec:Intro}
Arboviruses can spread quickly and cause major disease epidemics. Mosquitoes are vectors of some of the world’s most severe diseases such as malaria, dengue, Zika, Chikungunya, and West Nile Virus disease \cite{who_chikungunya, who_dengue, who_malaria, who_zika, who_wnv}. Different modelling approaches have been developed to simulate the abundance and seasonal dynamics of mosquito vectors and support disease risk prediction. They can be viewed in two broad categories: mathematical and statistical models. Mathematical models rely on laboratory and field data for the parameterisation of key life history traits such as the development and mortality rates of different stages in the mosquito life cycle \cite{tran2013, erickson2010, cailly2012, marini2019, qiang2022, petric2020}. On the other hand, statistical models use correlative and other machine learning (ML) techniques to infer the relationship between vector abundance and a set of abiotic factors \cite{otero2006, chuang2012, edwards2021, oluwagbemi2013, guisan2005, dare2022, dare2023}. These models typically require multi-year time-series of mosquito surveillance data (derived from labour intensive longitudinal field studies) to produce accurate outputs, and are often subject to various sources of variability that could lead to biased results. 

In this paper we explore the feasibility of using Physics Informed Neural Networks (PINNs) trained on an ordinary differential equation (ODE) mosquito population dynamics model to bridge the gap between conventional mathematical modelling and data-science approaches while conserving the physical and biological constraints and consistencies that govern these systems and processes.

\subsection*{Physics Informed Neural Networks}
Recent advancements in computational capabilities and the exponential growth in data availability have made data-driven analytics one of the predominant strategies in both research and practical applications.
Deep learning using different forms of neural networks is central to this development and has been extensively applied across various domains such as computer vision \cite{NEURIPS2021_cba0a4ee, Quach2023}, natural language processing \cite{NIPS2017_3f5ee243, tran2021hierarchical}, genomic prediction in plants \cite{Azodi2019} and finance \cite{NEURIPS2022_0bf54b80, 10.1007/978-3-031-26438-2_28}.
In data-driven approaches, neural networks are trained to minimize discrepancies between model predictions and observed data.
However, this purely data-driven approach has a number of limitations, including poor interpretability \cite{SAEED2023110273}, poor out-of-distribution generalization \cite{Yu2024ASO}, and the requirement for substantial amounts of training data \cite{Karniadakis2021}.

Physics-informed neural networks (PINNs) \cite{RAISSI2019686} have emerged as an alternative for scenarios where data is governed by underlying physical laws expressed through differential equations. 
This approach integrates domain-specific knowledge into machine learning by incorporating physical laws as additional objective loss functions alongside the traditional data-fitting loss functions.
This multi-task optimization strategy not only attempts to align with observational data but also to approximate the governing differential equations.
As a result, PINNs adhere to physical laws and thereby, enhance model generalizability and also uncover latent patterns within empirical data. 
This type of framework also facilitates an uncomplicated solution to both forward and inverse problems, where we can simultaneously learn the system state (forward problem) and the system's parameters (inverse problem).
Successful deployments of PINNs have been demonstrated across various fields, as surveyed in \cite{Karniadakis2021, Banerjee2023PhysicsInformedCV, Cai2021PhysicsinformedNN, hao2023physicsinformed, Banerjee2023ASO, 10.1371/journal.pcbi.1007575}.

However, despite this potential, PINN deployments continue to face significant challenges, particularly in training models that involve multi-scale and stiff solutions \cite{Karniadakis2021, WANG2021113938, Wang2020UnderstandingAM}.
In recent years, numerous advancements have been made to enhance the foundational framework initially proposed in \cite{RAISSI2019686}.
These enhancements include the development of innovative neural network architectures \cite{WANG2021113938, Wang2020UnderstandingAM, Gao_2021, REN2022114399} and novel adaptive activation functions \cite{JAGTAP2020109136, Jagtap_2020}.
Several studies have focused on optimizing the multi-task training process by adaptively adjusting the weights of different loss components \cite{Wang2020UnderstandingAM, Maddu_2022, WANG2022110768}.
Others have explored modifications in the distribution of collocation points \cite{doi:10.1137/19M1274067, WU2023115671, https://doi.org/10.1111/mice.12685, TANG2023111868}.
Additionally, some researchers have adopted a sequential learning approach, where training is conducted on one subdomain at a time before progressing to the next \cite{CiCP-29-930, krishnapriyan2021characterizing, MATTEY2022114474}, thereby preserving the causality within the system \cite{Wang2022RespectingCI}.
Moreover, there are efforts where the input domain is divided into smaller subdomains with PINNs  trained separately on each subdomain, significantly enhancing the model's convergence and accuracy \cite{JAGTAP2020113028, jagtap2020extended, Moseley2021FiniteBP}.

\subsection*{Contribution and Paper Structure}
While the majority of research on PINNs has concentrated on enhancing techniques for the effective training of Partial Differential Equations (PDEs), there has been less focus on customizing these methods for multi-variate, multi-equation Ordinary Differential Equation (ODE) systems.
In this study, we explore the applicability and effectiveness of existing PINN methodologies to ODE systems. 
We introduce specific adjustments designed to optimize the training of ODE systems, including the individual normalization and loss weight balancing tailored for each variable and equation involved in the system.
Moreover, we implement two additional steps in the training process to enhance both the effectiveness and accuracy of the models, and to simplify domain decomposition for practical application scenarios. 
We evaluate our modified framework using the Lorenz system, a classical model in dynamical systems theory, to demonstrate its applicability and value. 
Finally, we apply our approach to the modelling of mosquito population dynamics to validate its effectiveness in a practical biological context.

The contributions of this research can be articulated as follows:
\begin{itemize}
    \item The development of a systematic framework for training physics-informed neural networks (PINNs) on real-world ordinary differential equation (ODE) systems.  
    This framework incorporates a range of customized techniques, including ODE normalization, gradient balancing, causal training, and domain decomposition, to address common challenges in training PINNs with ODE systems.
    
    \item Our method includes a comprehensive normalization not only for inputs and outputs but also ODEs, ensuring more stable and accurate training processes in multi-variate multi-equation systems. In this respect, we implement an adaptive re-weighting of loss functions that individually adjusts the ODE loss weights, thus ensuring balanced training across physics constraints and data loss.
    Model training is further enhanced by incorporating a 3-phase progressive learning approach that respects temporal causality. This begins with data fitting for initialization, followed by cumulative training across sub-domains of increasing size, and informed neural networks guide mechanistic modelling from sparse experimental data finally tuning across the entire domain.
    In addition, we simplify domain decomposition, enforcing equality of boundary values at the interface between sub-domains and as a result, avoid extensive calculations to achieve continuity in high-order derivatives.
    
    \item A robust 2-step validation is carried out, firstly through an ablation study involving the Lorenz system and secondly, using mosquito population dynamical modelling, to validate the effectiveness of our approach and demonstrate its potential for continued study.
\end{itemize}

The remainder of this paper is structured as follows: 
section \nameref{sec:related_work} reviews the literature relevant to our study; 
section \nameref{sec:methodology} introduces our PINN framework as applied to a system of ODEs and including our proposed extensions;
in section \nameref{sec:lorenz_system}, an ablation study using the Lorenz system is used to evaluate the effectiveness of each component in our framework;
section \nameref{sec:mosquito} examines the wider impact and applicability of our approach by modelling mosquito populations;
and finally in section \nameref{sec:discussion}, we conclude with some limitations of our current methods with suggestions for future work in this important research area.

%
%
%
%
\section*{Related Work}
\label{sec:related_work}

\subsection*{Mosquito Population Dynamics Modelling}
The main methods that are currently being used in mathematical modelling of mosquito dynamics are linear and nonlinear systems of coupled ordinary (ODE) \cite{tran2013, erickson2010, cailly2012, marini2019, petric2020} or delayed differential equations (DDE) \cite{qiang2022, huang2020, song2020}. These models are compartmental, stage-structured models which divide the population into sub-groups corresponding to the developmental stages of the mosquito vector. They typically include at least four compartments separating the aquatic or immature stages (egg, larva and pupa) from the airborne adult stages, but are usually more complex, including additional adult stages to accurately simulate resting, feeding and gestating females, as well as density dependent competition within the immature stages \cite{jucht2015}. The rate at which individuals progress from one stage to the other is simulated by the species-specific development and mortality rates, which depend on micrometeorological variables including air temperature, precipitation and relative humidity, as well as a complex set of interactions between the individuals of the same and competing species, breeding site availability etc. \cite{roiz2011, carrieri2014, groen2017, marini2019}. These dependencies introduce temporal constraints to the system in terms of “stiffness” which has an effect on the overall stability and the numerical integration \cite{kass1997introduction,desolve2012, dahlquist2003}. The magnitude of the stiff problem is defined by the time scale and variability of the driving biotic and abiotic processes. This determines the choice of the time-differencing scheme used to solve the equations, number of steps, local accuracy and length of the numerical integration \cite{desolve2012, dahlquist2003}. A common issue in current approaches is the lack of experimental data for the accurate calibration of the development and mortality parameters, as well as applicability to locations with different ecoclimatic settings \cite{erguler2022dyn}. Although there has been recent research on the use of physics-informed neural networks in the field of biological systems \cite{Daryakenari2024, Lagergren2020}, this is, to the best of our knowledge, the first paper to investigate the feasibility of applying PINNs to an ODE mosquito population dynamics model. In this study, we aim to provide a first step toward an integrated deep learning vector population dynamic modelling framework.



\subsection*{Normalization}
Normalization is a crucial yet often overlooked step in the training of PINNs.
In \cite{Moseley2021FiniteBP}, researchers employed a strategy that involved dividing the input domains and applying individual input normalization alongside a unified global output normalization within their model computations.
In \cite{wang2023experts}, the authors suggested not only normalizing the inputs and outputs of the models but also non-dimensionalizing the differential equations integrated into the objective functions. 
In both \cite{Lagergren2020, 10.1371/journal.pcbi.1007575} and \cite{Daryakenari2024}, both approaches added input- and output- scaling layers that multiply the inputs and outputs with their average magnitudes. However, this is carried out at the model level and thus, still affects the objective functions and training efficiency.
Despite recognizing the advantages of these approaches, there remains a lack of a systematic methodology for normalization, particularly in the context of dynamical systems characterized by multivariate system states and numerous differential equations.
To address this in our research, we developed a normalization procedure that applies the MIN-MAX scheme to both the inputs and outputs of the neural networks, while appropriately transforming the system of ODEs. This procedure aids not only in aligning with the assumptions of model initialization but also in addressing the challenges associated with multi-scale and stiff issues commonly encountered in training PINNs.

\subsection*{Loss Re-weighting}
PINNs operate as a multi-task learning framework that incorporates distinct losses for data fidelity and adherence to physical laws. The different scaling and convergence rates of these losses can lead to imbalances, potentially direct the model towards incorrect solutions as one objective may disproportionately influence the training process.
A common solution is the re-weighting of losses to achieve a more balanced training.
\cite{Wang2020UnderstandingAM} demonstrated that one of the primary training pathologies in PINNs is the imbalance in gradients propagated from the different losses.
These authors proposed an adaptive method that adjusts the weights of the losses based on the ratio between the maximum gradient magnitude of the physics task loss and the mean gradient magnitude of the data task loss, with respect to the model parameters. The authors in 
\cite{wang2023experts} used a similar approach utilizing the ratio of the $L_2$-norm of the gradients instead while in \cite{Maddu_2022}, researchers opted to balance the variances of the gradients instead. In \cite{WANG2022110768}, they applied the Neural Tangent Kernel to demonstrate that losses from physical laws converge substantially faster than those from initial or boundary conditions, proposing an algorithm to equalize the convergence rates by monitoring the kernels of the losses.
We differ to the above approaches, in that while we adopt the strategy proposed by \cite{Wang2020UnderstandingAM}, we did so with several modifications.
Similar to \cite{10.1371/journal.pcbi.1007575}, we set the weight for the data loss at a fixed value of 1.0 and adjusted the weights for the physics law losses, which exhibited more significant variability.
And we assigned individual weights to different differential equations, providing the flexibility needed to accommodate the diverse scales and behaviors of these equations.

\subsection*{Collocation points}
Collocation points (residual points), where physics constraints are minimized, are traditionally selected uniformly at random across the domain.
However, this uniform approach may not be optimal for systems exhibiting steep derivatives.
A method known as residual-based adaptive refinement (RAR) proposed by \cite{doi:10.1137/19M1274067} enhances this process by adding new collocation points with the highest differential equation residuals every few iterations, enabling models to focus adaptively on the most challenging areas during training.
In \cite{WU2023115671}, the authors adopt a similar strategy by sampling collocation points based on a probability distribution proportional to these residuals. 
Elsewhere \cite{https://doi.org/10.1111/mice.12685}, researchers used importance sampling technique to derive a distribution proportional to the 2-norm of the gradient of the loss function, approximated by the loss value to lower computational demands. In
\cite{TANG2023111868}, they also derived an improved collocation point distribution but use a generative deep learning model to approximate the distribution.

In \cite{krishnapriyan2021characterizing}, the authors implemented a sequential training approach, dividing the input domain into subdomains and using predictions from one as initial conditions for the next. 
Conversely, researchers in \cite{MATTEY2022114474} leverage predictions from all prior subdomains to enable a single global approximation network.
A different approach is presented in \cite{CiCP-29-930, baty2023solving} where the authors introduce a progressive learning approach with residual points drawn uniformly from a dynamically expanding subdomain, starting from a single point and growing to cover the desired domain. This way, the training process starts by solving the ODEs at earlier in time before moving to what happens later.  This technique respects the time causality essential for accurately predicting dynamical systems' evolution.
The authors in \cite{Wang2022RespectingCI} also emphasize time causality by weighting the residuals to prioritize earlier time points.

We build on the methods presented in \cite{CiCP-29-930, baty2023solving} but crucially add two new steps: one for initial data fitting to improve model initialization (such as \cite{10.1371/journal.pcbi.1007575, Daryakenari2024}) and include a final step which tunes across the entire domain.

\subsection*{Domain Decomposition}
When dealing with extremely large input domains, convergence in training PINNs can be particularly challenging.
A domain decomposition approach addresses this by dividing the domain into smaller subdomains and training PINNs for each.
To maintain solution continuity and smoothness across these subdomains, additional objective functions, known as interface losses are integrated into the optimization.
In \cite{JAGTAP2020113028}, the authors introduce two interface conditions: one ensures the alignment of solution values at the interfaces, while the other enforces the consistency of conservation laws across these boundaries. For inverse problems, these conditions also extend to parameter values at the interfaces. 
The authors in \cite{jagtap2020extended} further expanded on these interface conditions to accommodate arbitrary differential systems by ensuring the continuity of the differential equations at the interfaces. In \cite{Moseley2021FiniteBP, 10.1007/978-3-030-63393-6_2}, they adopt an implicit approach by employing a gating function.
For real-world applications where a high degree of smoothness in the solution is negligible, we simplify the approach by \cite{jagtap2020extended}. Instead, we opt to implement only the value enforcement at the interfaces, which serves both as the initial condition and as a means to ensure continuity across the subdomains.


%
%
%
%
\section*{Methods}
\label{sec:methodology}
In this section, we formulate the problem and over a number of steps, present our proposed methodology which creates a solution for problems employing Ordinary Differential Equations (ODEs).

\subsection*{PINN Structure}

Consider $u(t) = \left(u^{(1)}, u^{(2)}, \dots, u^{(V)} \right)$ as a $V$-dimensional vector representing the state of a dynamical system at any given time $t$, where $t$ ranges from 0 to $T$. This dynamical system $u$ is governed by a set of $F$ ODEs as shown in Eq~\eqref{eq:ode_system}, where \ldots

\begin{align}
\frac{d u}{dt} = f^{(i)}\left(t, u, \theta\right), \quad i=1, 2, \dots, F.
\label{eq:ode_system}
\end{align}

In these equations, the system's behavior over time is shaped by a set of $P$ parameters $\theta=\left(\theta^{(1)}, \theta^{(2)}, \dots, \theta^{(P)} \right)$, which might or might not be known in advance. The functions $f^{(i)}$ are known functions defining the system's dynamics. Suppose that we have some observations of the system at different times, $\mathcal{D}_u = \left\{ (t_1, u_1), (t_2, u_2), \dots, (t_j, u_j), \dots \right\}$. Our objective is to find a solution $u$ and possibly $\theta$ that simultaneously matches these observations and is consistent with the ODEs.

In the standard PINN framework \cite{RAISSI2019686}, a neural network $U$ (parameterized by ${W_U}$), is used to approximate the solution $u$. This network, as a function defined in $[0, T]$, tries to estimate $u$ at any given time $t$, with $W_U$ being the trainable parameters of the network. 
In inverse problems where a few or all the parameters $\theta$ is not available, we can use a neural network, $\Theta^{(l)}$, to predict the unknown values $\theta_l$.
For the sake of simplicity, we denote the set of all parameters, including the known or ones to be learnt by neural networks, as $\Theta$ and regard it as neural networks.
The parameters $\Theta$, parameterized by $W_\Theta$, are also defined as functions in the domain $[0, T]$. 
If we let $W=\{W_U, W_\Theta\}$, then the task now becomes an optimization problem when determining the parameters $W$ as shown in Eq~\eqref{eq:W}, which aims to minimize the multi-task objective function defined in Eq~\eqref{eq:total_loss_fn}.

\begin{align}
    W = \text{argmin}_W \mathcal{L}
    \label{eq:W}
\end{align}

\begin{align}
    \mathcal{L} = \mathcal{L}_{data} + \frac{1}{F} \sum_{i=1}^F \lambda_i \mathcal{L}_{f^{(i)}}
    \label{eq:total_loss_fn}
\end{align}

where

\begin{align}
    \mathcal{L}_{\text{data}} &= \frac{1}{N_u} \sum_{(t_i, u_i) \in \mathcal{D}_u} (U(t_i) - u_i)^2 
    \label{eq:loss_data}
\end{align}

\begin{align}
    \mathcal{L}_{f^{(i)}} &= \frac{1}{N_f} \sum_j \left|\frac{d U}{dt} - f^{(i)}\left(t_j, U(t_j),\Theta(t_j)\right)\right|^2 
    \label{eq:loss_ode}
\end{align}

Where $U(t_j), \Theta(t_j)$ are the output values of the neural networks valuated as $t_j$. By minimizing the objective $\mathcal{L}_{\text{data}}$, we decrease the discrepancy between the network predictions and the observed data. 
Likewise, by minimizing the residuals  with $\mathcal{L}_{f^{(i)}}$, the neural network $U$ aligns with the differential equations at low errors. 
Here, $\lambda_i$ are balancing factors between fitting to the data and adhering to the dynamics, and $N_f$ is the number of residual points randomly sampled from a distribution $\mu$, typically uniform, in the domain $[0, T]$. 
We resample the residual points every step to ensure the losses are minimized everywhere in the entire period.
Overall, by minimizing this overall loss function $\mathcal{L}$, the network can fit observed data while also approximately follows the dynamic rules at the same time.

The loss function is minimized using gradient-based algorithms. The methods iteratively update the parameters $W$ in the directions of reducing the loss function, based on its gradients with respect to the parameters, namely, $\nabla_W \mathcal{L}$. Specifically, the updates are performed according to the following rule
\begin{align}
    W_{\text{new}} = W_{\text{old}} - \eta \nabla_W \mathcal{L} = W_{\text{old}} - \eta  \left( \nabla_W \mathcal{L}_{data} + \sum_{i=0}^F \lambda_i \nabla_W \mathcal{L}_{f^{(i)}} \right)
    \label{eq:update_rule}
\end{align}
where $\eta$ is the learning rate, which dictates how big of a step to take in the direction opposite to the gradient. To carry out this optimization, the Adam algorithm \cite{Kingma2014AdamAM}, a variant of the gradient descent algorithm that has been widely used on training PINNs, is used.
The calculations of gradients, whether it is the network $U$ with respect to time $t$ or the loss function $\mathcal{L}$ with respect to $W$, can rely on automatic differentiation supported by popular deep learning frameworks such as Pytorch \cite{Paszke2019PyTorchAI}, Tensorflow \cite{abadi2016tensorflow} or JAX \cite{jax2018github}.

According to the universal approximation theorem \cite{LESHNO1993861}, multi-layer perceptrons (MLP) \cite{haykin1994neural} can approximate any continuous function on a given domain provided the MLP has sufficient complexity, as represented by the number of hidden layers and parameters in those hidden layers. It is shown that challenges in PINN may not be due to the capabilities of MLP \cite{krishnapriyan2021characterizing}. For an MLP with $L$ layers, the mathematical formulation that describes its operation, layer by layer, is presented in Eq 
\eqref{eq:layers} where $x$ is the input to the neural network, $h^{(l)}$ is the hidden state at layer $l$, $W^{(l)}$ and $b^{(l)}$ are the parameters of the layer $l$, and $\sigma$ is the activation function which provides non-linearity for the model.

\begin{align*}
\label{eq:layers}
h^{(0)} &= x \\
h^{(l)} &= \sigma (W^{(l)} h^{(l-1)} + b^{(l)}), l=1,\dots,L-1 \numberthis \\
h^{(L)} &= W^{(L)} h^{(L-1)} + b^{(L)}
\end{align*}

The GELU activation function \cite{hendrycks2023gaussian} is employed for its smooth properties which is essential for differential problems, offering an advantage over the ReLU function. For the initialization of the MLP's weights and biases, the Glorot scheme \cite{glorot10a} is utilized.

The "vanilla" PINN framework described above can approximate well when the ODE systems are relatively simple. However with more complex systems, especially ones that exhibit extreme stiffness, chaotic and multi-scale behavior, the basic setup tends to have difficulties in converging to a satisfactory local minimum. For this reason, we now proceed to describing our approaches to aid PINN training for both forward and inverse problem involved ODE systems.

\subsection*{ODE Normalization}
Under normal circumstances, NNs' output is in the range [-1,1] and so for output values outside this range, data normalization plays a crucial role in the machine learning workflow, ensuring that both input and output variables remain within a reasonable range to enhance model convergence and accuracy. In the PINN context, this normalization process requires careful consideration, as any transformation results in modification to the corresponding ODE systems. In \cite{10.1371/journal.pcbi.1007575, Moseley2021FiniteBP}, the authors consider normalization and de-normalization as parts of the overall model. However, this type of approach retains the original variable scales of the variables, potentially leading to significant imbalances in the objective function. \cite{wang2023experts} suggests normalizing the differential equations as well but its generalization is omitted. In this paper, we propose a systematic approach to do normalizations using PINNs involving ODEs. In particular, we utilize the MIN-MAX normalization scheme for both input and output variables, i.e. $t$, $u$ and $\theta$. To normalize the input variable, i.e. time $t$, we employ the transformation shown in Eq~\eqref{eq:ttrans}, where $[T_{\min}, T_{\max}]$ represents the time domain in which the models are trained. This will have the effect of normalizing the time variable $t$ to the range of $[-1, 1]$.

\begin{align}
\label{eq:ttrans}
    t' = 2 \cdot \frac{t-T_{\min}}{T_{\max} - T_{\min}} -1
\end{align}

To normalize the outputs from the neural networks, we define $\mathfrak{L}_u, \mathfrak{U}_u, \mathfrak{L}_\theta$, and $\mathfrak{U}_\theta$ as the lower and upper bounds for $u$ and $\theta$, respectively. When the difference between the lower and upper bounds is small, we adjust the bounds around the mean as $\mathfrak{L} = \min(\mathfrak{L}, \mathfrak{M} - 1)$ and $\mathfrak{U} = \min(\mathfrak{U}, \mathfrak{M} + 1)$ where $\mathfrak{M} = \frac{\mathfrak{L} + \mathfrak{U}}{2}$ is the middle point. It is important to note that these bounds are specific to each dimension in the dynamical system. These bounds can be estimated through collected data, inferred from domain knowledge, or estimated through approximate simulations when data is scarce. The normalization of the variables $u$ and $\theta$ at a normalized time $t'$ is then obtained using Eqs~\eqref{eq:u_norm} and \eqref{eq:theta_norm} respectively.

\begin{align}
    u'(t') &= \frac{u(t') - \mathfrak{L}_u}{\mathfrak{U}_u - \mathfrak{L}_u} \cdot 2 - 1 \Longleftrightarrow u(t') = (u'(t') + 1) \frac{\mathfrak{U}_u - \mathfrak{L}_u}{2} + \mathfrak{L}_u
    \label{eq:u_norm} \\
    \theta'(t') &= \frac{\theta(t') - \mathfrak{L}_\theta}{\mathfrak{U}_\theta - \mathfrak{L}_\theta} \cdot 2 - 1 \Longleftrightarrow 
    \theta(t') = (\theta'(t') + 1) \frac{\mathfrak{U}_\theta - \mathfrak{L}_\theta}{2} + \mathfrak{L}_\theta \label{eq:theta_norm}
\end{align}

Following the normalization above, the transformed inputs and outputs, $t'$, $u'$ and $\theta'$, now vary within the range [-1, 1].
Thus, instead of having neural networks approximating $u$ and $\theta$, it is instead more beneficial to use $U$ and $\Theta$ as surrogate models for $u'$ and $\theta'$. Following the normalization transformation in Eqs~\eqref{eq:u_norm} and \eqref{eq:theta_norm}, it is necessary to adapt the loss functions accordingly. When computing $\frac{d u'}{ d t}$ and applying Eq~\eqref{eq:ode_system}, we arrive at the result shown in Eq~\eqref{eq:dudt}.

\begin{align}
    \label{eq:dudt}
    \frac{d u'}{ d t} = \frac{2}{\mathfrak{U}_u - \mathfrak{L}_u} \frac{d u}{ dt} = \frac{2}{\mathfrak{U}_u - \mathfrak{L}_u} f^{(i)}\left(t, u, \theta\right)
\end{align}

Instead of using objective functions \eqref{eq:loss_data} and \eqref{eq:loss_ode}, we now consider the alternative Eqs~\eqref{eq:loss_data_normed} and \eqref{eq:loss_ode_normed} as these will provide re-scaling of the losses to the magnitude order of $u$ and $\frac{d u}{ dt}$, which are more similar to each other than the original objective functions are.   

\begin{align}
    \mathcal{L}_{\text{data}} &= \frac{1}{N_u} \sum_{j} \left(U(t_j') - \frac{u_j - \mathfrak{L}_u}{\mathfrak{U}_u - \mathfrak{L}_u} \cdot 2 - 1 \right)^2 \label{eq:loss_data_normed} \\
    \mathcal{L}_{f^{(i)}} &= \frac{1}{N_f} \sum_j \left| \frac{d U}{dt} - \frac{2}{\mathfrak{U}_u - \mathfrak{L}_u} f^{(i)}\left(t_j, (U(t_j')+1) \frac{\mathfrak{U}_u - \mathfrak{L}_u}{2} + \mathfrak{L}_u, (\Theta(t_j')+1) \frac{\mathfrak{U}_\theta - \mathfrak{L}_\theta}{2} + \mathfrak{L}_\theta \right)\right|^2 
    \label{eq:loss_ode_normed}
\end{align}

The modifications to the objective functions, as detailed above, not only trying to balance the data loss and ODE terms, and also across the ODE components.
This re-scaling is particularly beneficial in scenarios where variables exhibit significant differences in scale, arising from their inherent characteristics or the units used for measurement.
By normalizing these variables, we prevent any single variable from dominating others, balancing the impact of each terms the training process and hence enhancing the convergence. 
This approach is also aligned with the assumptions of Glorot initialization for neural networks \cite{glorot10a}. 
However, despite these normalizations, $\frac{d U}{d t}$ might not be completely re-scaled and could still reflect the inherent stiffness of the system.
Further measures need to be taken, such as domain decomposition and weight re-balancing discussed in other subsections.

\subsection*{Gradient Balancing}

As this is a multi-task problem, we have different objective functions (e.g. Eq     \eqref{eq:total_loss_fn}) as we are measuring different constraints. Previous research  \cite{Wang2020UnderstandingAM} highlighted that one point of failure in PINN training is the imbalance in gradients $\nabla_W \mathcal{L}_{data}$ and $\nabla_W \mathcal{L}_{f^{(i)}}$ in the update rule \eqref{eq:update_rule}.
It is observed that the differential equation residual loss dominates the overall loss due to the \emph{stiffness} of the system. 
This makes the model prioritise optimizing the ODE constraint over matching the initial, boundary condition or data observations, leading the model to converge to a trivial solution, i.e. the \emph{null solution},  violating conditions for the data. 
We address this issue by adopting a similar approach to \cite{Wang2020UnderstandingAM} where 
balancing adjusts the weights $\lambda$ in the objective function based on statistics from the gradients $\nabla_W \mathcal{L}_{data}$ and $\nabla_W \mathcal{L}_{f^{(i)}}$.
We extend this further to ODE systems by separately and individually assigning and adjusting the weights to each and every differential equation. 
In objective functions \eqref{eq:total_loss_fn} and \eqref{eq:update_rule}, the weights for data are set to 1 with component weights $\lambda_i$ each adjusted individually, to ensure that every equation in the system is given equal \emph{importance} while aligning with the \emph{importance} given to data.
This extension to previous work is crucial because the scale and complexity of each equation are different and require different treatment during training.

\begin{algorithm}[H]
\caption{gradient\_balancing() }
\label{alg:gradient_balancing}
\begin{algorithmic}

\Require \text{step} $\gets$ current training step; $\alpha$ smoothing factor; updates are made every $N$ steps
\Ensure re-calculating $\lambda_i$ such that gradients from different loss terms are balanced 

\If{$\text{step} = 0$}
    \State Initialize $\lambda_i \gets 1, \forall i = 1, \dots, F$
\EndIf
\If{step $\bmod N$ = 0}
\State Compute $\hat{\lambda}_i$ by 
\begin{align}
    \hat{\lambda}_i \gets \frac{\overline{\left|\nabla_W \mathcal{L}_{data}\right|}}{\max \left\{ \left| \nabla_W \mathcal{L}_{f^{(i)}} \right| \right\}}, i= 1,\dots,F
\end{align}
where $\overline{\left|\nabla_W \mathcal{L}_{data}\right|}$ is the average of the absolute gradients over all model parameters $W$.
\State Adjust the weights $\hat{\lambda}_i$ by
\begin{align}
    \lambda_i \gets \alpha \cdot \lambda_i + (1-\alpha) \hat{\lambda}_i, i = 1, \dots, F \label{eq:update_lambda}
\end{align}
\EndIf

\end{algorithmic}
\end{algorithm}

This gradient-balancing algorithm is described in Algorithm \ref{alg:gradient_balancing}, where all weights $\lambda_i$ are initialized to 1 and are updated every $N$ steps.
We compute the weight $\hat{\lambda}_i$ by calculating the ratio between the mean of absolute values of gradient $\nabla_W \mathcal{L}_{data}$ and the maximum of the absolute values for the gradients of $\nabla_W \mathcal{L}_{f^{(i)}}$.
Due to the stochastic nature of gradient descent, the weight from these calculations can be highly volatile and thus, we update the weights $\lambda_i$ using the moving average formula in Eq~\eqref{eq:update_lambda}.
The recommended values for hyper-parameters $\alpha$ and $N$ in the original study \cite{Wang2020UnderstandingAM} are $\alpha=[0.5, 0.9]$. However, we set $N=100$ and tune $\alpha$ in extreme cases $\alpha=0.99, 0.9, 0.5$, or set $N=1$ with $\alpha=0$.

\subsection*{Causal Training}
When training PINNs, it is important to consider the \emph{order} of causal effect \cite{Wang2022RespectingCI}, especially when dealing with problems where data is quite sparse.
Usually, PINN models are trained to follow differential equations at every point in the input domain at the same time. 
However a problem can arise where the model starts to follow these rules at later values for $t$ but it has properly adhered to the rule at an earlier point $t$.
This discrepancy leads to a situation where efforts to conform to ODEs at a one point cause violations at other points during the learning process.
Furthermore, if the penalty for not following ODEs at the earlier point outweighs the fitting at the later, the model may get stuck in a local minimum, unable to satisfy ODEs over the entire period anymore.
To solve this, a \emph{causal} approach to training is recommended. In a causal training, the task of meeting the data conditions is given priority while ensuring that the model complies with ODEs at the earlier values $t$ before attempting compliance later values $t$.

In this study, we divide the training process into three phases: data fitting, progressive causal training, and final tuning.
\begin{itemize}
    \item First, in the data fitting phase, we focus on making the model match the data conditions but training only with the data loss term $\mathcal{L}_{\text{data}}$.
This helps set a good starting point for the model so that later, it can follow the rules of differential equations more easily.
    \item In the progressive causal training phase, we take the \emph{growing-interval} approach used in \cite{baty2023solving}. 
We \emph{gradually} teach the models to follow the differential equations starting from a small interval and slowly covering more domain as the training progresses.
Both data loss term and ODE loss terms are included but the ODE residual points for $\mathcal{L}_{f^{(i)}}$ are drawn from a growing interval.
If $N_2$ is the number of update steps in the second phase, then at the $n_2$-th step, residual points are uniformly drawn from the interval $\left[ 0, \frac{n_2}{N_2} \cdot T \right]$.
    \item In the final tuning phase, we train the model using both data loss and ODE losses, with the ODE residual points drawn at random with uniform distribution across the entire domain. 
    The goal here is to refine and improve the solutions the model has learned so far, ensuring it follows the differential equations more accurately over the entire period.
\end{itemize}

The training process mainly happens during the second and third stages, with the first stage being the shortest. This is because neural networks can quickly learn to fit the given data too well. We usually limit the first phase, where the model learns to match the initial data, to between 5,000 and 10,000 steps.
In the progressive causal phase, the model gradually learns to solve the differential equations over an expanding area. This phase takes longer because the model must successfully reduce its errors to a required level before it can extend its learning to new areas.
For systems that do not reliably converge, the number of steps for this phase should be sufficiently high. 
We set this phase to last between 50,000 and 100,000 steps to ensure thorough learning.
Then, in the final tuning phase, the early stopping approach is adopted. 
The model's performance is checked every 1,000 steps with an evaluation loss, where all weights $\lambda$s are set to 1.0. 
If no improvement is recorded in the model's performance after several checks, the training ends.
This ensures the model is as accurate as possible without time spent on unnecessary training.

\subsection*{Domain Decomposition}
One of the issues when working with PINNs is that the time series may be too large. Domain decomposition is a effective way to train PINN when the solution is overly complex for a particular time interval or the target input domain is too large, which is often the case in real world applications. 
The method is particularly useful in the extrapolation forward problem where there are little or no data in the period of interest, or the data condition (i.e. initial or boundary condition) needs to rely on predictions from some other domain. 
In such cases, the shape of the solution relies heavily on training the models to fit the ODE constrains, which is a difficult task due to issues such as causal effect violation \cite{Wang2022RespectingCI} or gradient imbalance \cite{Wang2020UnderstandingAM}.
The decomposition of domains reduces the domain the models are trained on and subsequently, reduces the complexity of the optimization task.

In particular, the approach divides the input domain into $S$ non-overlapped subdomains, $D_s = [T_{s-1}, T_s], s = 1, \dots, S$, with $T_1 = 0, T_S = T$. 
In each subdomain $D_s$, a neural network $U_s$ is defined, with the overall solution is defined in Eq~\eqref{eq:subdomain_u}. With this definition, the goal is to train $U_s$ to approximate the solution $u$ in the subdomain $D_s$. 

\begin{align}
    U(t) = \sum_s \mathbf{1}_{D_s}(t) \cdot U_s (t), \forall t \in [0, T]
    \label{eq:subdomain_u}
\end{align}

Let $D_s^* = [T_{s-1} - O, T_s + O]$ be the extended subdomain of $D_s$, $O$ be 50\% of overlapped size. We treat each subdomain $D_s^*$ as a separate PINN problem, where the model is trained using the framework described in section \nameref{sec:methodology}, including the normalization and gradient balancing measure. Starting from the first subdomain $D_0$, the model is trained with the initial condition data $\mathcal{D}_u$ provided by users. For subsequent subdomains $D^*_s$, the model is trained with data $\mathcal{D}_u$ generated by the previously trained model $U_{s-1}$ in the overlapped domain $[T_{s-1}, T_{s-1}+O]$. 
The volume of data generated could be unlimited, we set the number from 10 to 100, depending on the granularity and the continuity across subdomains required.
This process is repeated until the entire domain is covered. The final solution would be the combined predictions in Eq~\eqref{eq:subdomain_u}. 

This divide-and-conquer scheme offer several advantages in training PINNs: it reduces the complexity of the overall problem into many smaller less-complex problems; it reduces the explosion of gradients which has always been an important issue when training PINN; it allows optimized customization of models in each subdomain, enhancing the individual and overall convergence and accuracy.

In this study, we experiment with domain decomposition for the forward problem only. However, this approach could be expanded to inverse problem by adding similar interface conditions on  ODE-parameter neural networks.

%
%
%
%
\section*{Ablation Study with Lorenz System}
\label{sec:lorenz_system}
As different NN configurations or structures have been constructed for this task, we now evaluate the efficacy of the proposed PINN methodologies utilizing the Lorenz system, which consists of three coupled, nonlinear differential equations and is the common choice when evaluating ODEs \cite{Lorenz1963DeterministicNF}. This system is a classic example of chaotic behavior and is described in Eq~\eqref{eq:lorenz}. Here, $x$, $y$, and $z$ represent the state variables of the system at any given time $t$ while $\sigma$, $\rho$, and $\beta$ are physical parameters of the system.
We define two temporal domains for our experiments: a shorter duration with a maximum time of $T=2.0$ and a longer duration extending to $T=40.0$. 

\begin{align}
\label{eq:lorenz}
    \left\{
    \begin{array}{ll}
        \frac{dx}{dt} &= \sigma(y - x) \\
        \frac{dy}{dt} &= x(\rho - z) - y \\
        \frac{dz}{dt} &= xy - \beta z
    \end{array}
    \right.
\end{align}

The shorter duration is employed to assess the impact of techniques such as ODE Normalization, Gradient Balancing and Causal Training on the model's performance. 
Conversely, the longer duration is utilized to investigate the effectiveness of Domain Decomposition in managing extended temporal intervals.
For the generation of both training and ground-truth datasets, we employ a well-established numerical method \cite{doi:10.1137/0904010} for solving ODE systems, implemented in Python's Scipy library \cite{2020SciPy-NMeth}.

Normalization is a crucial preprocessing step in machine learning. 
In our study, we utilize ODE normalization as a baseline model and integrate various additional techniques to evaluate their impact on model performance.
We investigate the following models:

\begin{enumerate}
    \item $\text{OdePINN}_\text{orig}$: Original framework without additional techniques.
    \item $\text{OdePINN}_\text{baseline}$: ODE Normalization only. This model represents a baseline model for the more complex models 3-6 below.
    \item $\text{OdePINN}_\text{grad}$: ODE Normalization combined with Gradient Balancing.
    \item $\text{OdePINN}_\text{causal}$: ODE Normalization combined with Causal Training.
    \item $\text{OdePINN}_\text{grad+causal}$: ODE Normalization, Gradient Balancing, and Causal Training combination.
    \item $\text{OdePINN}_\text{grad+causal+domain}$: The model combines ODE Normalization, Gradient Balancing, Causal Training, and Domain Decomposition.
\end{enumerate}
Where a bespoke technique is not applied, conventional methods are used by default.
Specifically, in the absence of ODE normalization, input and output variables retain their original scale; without Gradient Balancing, the weights $\lambda_i$ are set to a fixed value of 1; without Causal Training, collocation points are uniformly sampled across the domain; and without Domain Decomposition, the model is trained on the entire domain of interest in a single training. In  this approach, experiments were conducted on both forward and inverse problems using the Lorenz system, setting a shorter time frame of $T=2.0$ to test the first five models. The sixth model $\text{OdePINN}_\text{grad+causal+domain}$ is assessed on the forward problem over a longer duration with $T=20.0$.

%
%

\subsection*{Forward Problem with T=2}
Now, we explore how the techniques described earlier work together, particularly on their capacity to extrapolate in forward problem scenarios using the Lorenz system. 
The data loss condition, $\mathcal{L}_{data}$, involves using only the initial condition as a single data point $(x, y, z) = (1, 1, 1)$ at $t=0$.
The physical parameters are set constant over time, as: $\sigma = 10$, $\rho = 28$, and $\beta = \frac{8}{3}$.
These parameters are all known to the framework.

The framework utilizes an MLP, $U$, to approximate the solution $u$. 
The architecture of $U$ comprises four hidden layers, each consisting of 100 units, with the GELU activation function applied in all hidden layers.
For Gradient Balancing, the hyperparameters $\alpha$ and $N$ are tuned across four settings: $(0.99, 100)$, $(0.9, 100)$, and $(0.0, 1)$. 
The configuration $(0.99, 100)$ is identified as the most effective through this tuning process.
In the absence of Gradient Balancing, $\alpha$ defaults to 0 and $N$ to None, maintaining the $\lambda_i$s at a constant value of 1.0 throughout the training.
Causal Training involves three phases: the first phase of 1,000 steps, followed by the second phase of 199,000 steps, and the final phase includes early stopping up to 100,000 steps.
Should Causal Training be deactivated, the model bypasses the initial two phases and proceeds directly to the last phase, extending training up to a maximum of 300,000 steps without the implementation of early stopping.

\begin{figure}[!ht]
    \centering
    \includegraphics[width=1.0\linewidth]{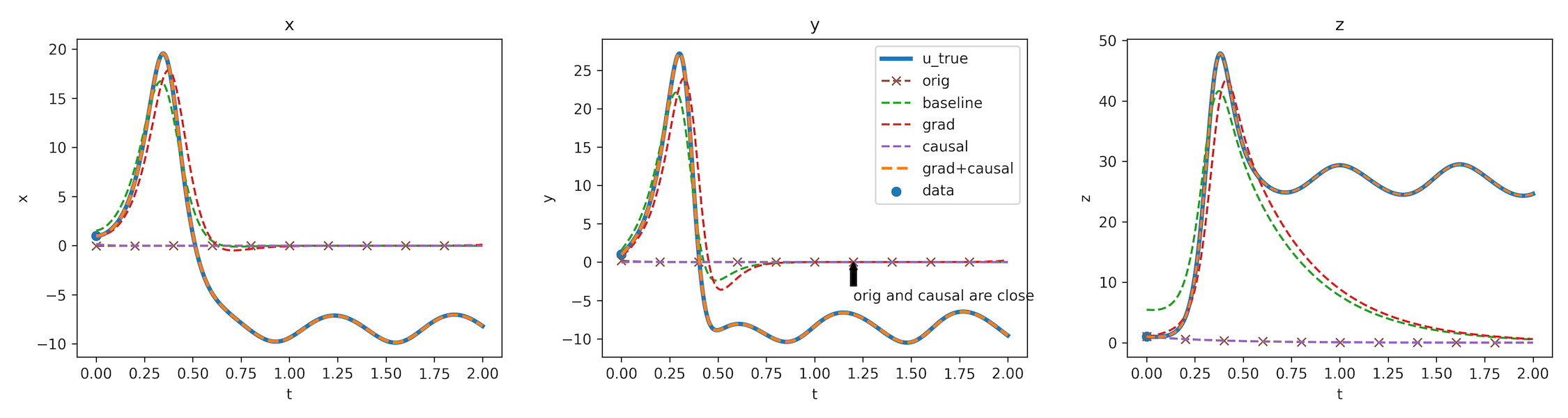}
    \caption{\textbf{Lorenz ODE system for the forward problem, with $U$ approximation of the system state.} The blue line u\_true represents the target for each of the 5 models used in the experiment. Graphs illustrate the performance across the x,y and z dimensions.}
    \label{fig:lorenz_forward_u_pred}
\end{figure}

\figurename~\ref{fig:lorenz_forward_u_pred} presents the solution approximations produced by the five models, starting with the poorest performing models.
\begin{itemize}
    \item $\text{OdePINN}_\text{orig}$ tends to converge towards the null solution due to the dominant factor of ODE loss.
    \item $\text{OdePINN}_\text{causal}$  is similar to the \textit{original} model, while capturing the system's dynamics effectively, deviates from the correct initial condition and subsequently converges to the trivial null solution.
    \item The baseline model manages to obtain the general shape of the solution but has difficulties in matching both the initial condition and the governing differential equations. 
    \item $\text{OdePINN}_\text{grad}$ achieves a better alignment with the initial condition but it fails to sufficiently satisfy the physics constraints, leading to inaccurate approximation of the correct solution.
    \item $\text{OdePINN}_\text{grad+causal}$ demonstrates a close convergence to the true solution, showing superior performance to the other models.
\end{itemize}

\begin{figure} [!ht]
  \centering
    \includegraphics[width=1.0\linewidth]{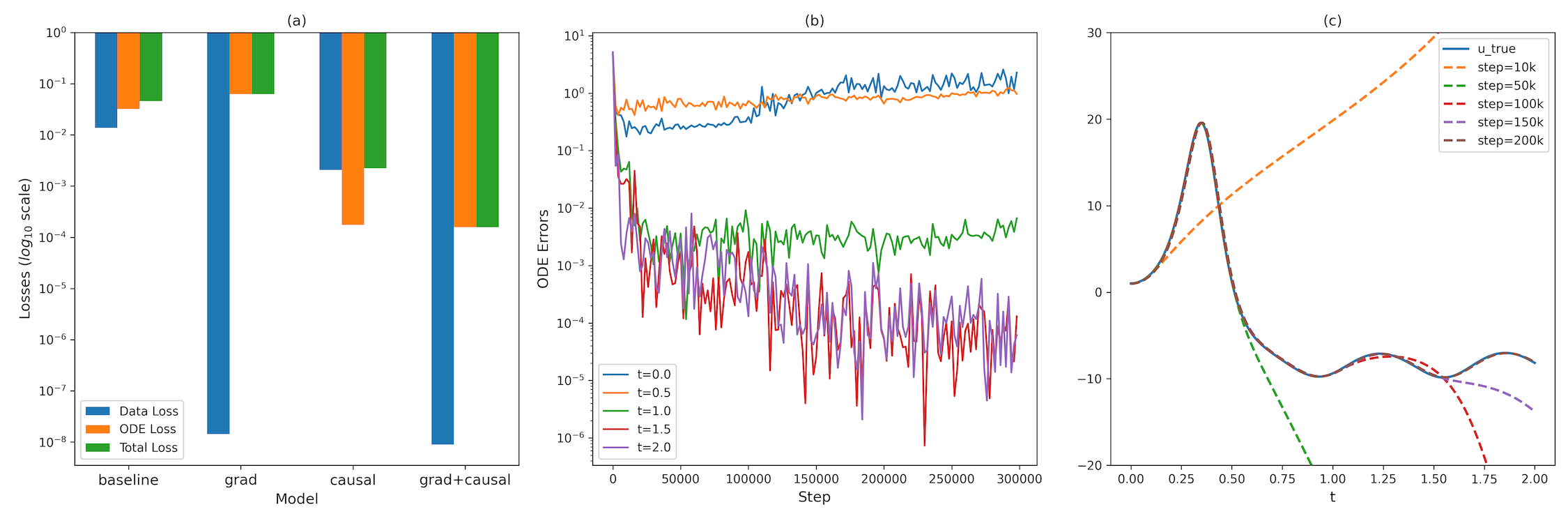}
  \caption{\textbf{Loss analysis of OdePINN framework with Lorenz system, $t \in [0, 2.0]$}. (a) Loss Terms across final models selecting through Early Stopping;
  the bars are presented in a logscale pointing downwards, with the lower the value the better the performance.
  (b) ODE errors of the $\text{OdePINN}_\text{grad}$ model at different time $t$ during the training, motivating the need for Causal Training;
  (c) The $x$-value Approximation Solution $U$ of the model $\text{OdePINN}_\text{grad+causal}$ at different steps during the training.
  }
  \label{fig:lorenz_forward_analysis}
  
  
\end{figure}

In summary, $\text{OdePINN}_\text{grad+causal}$ successfully integrates both methodologies, ensuring adherence to the initial condition and minimizing ODE loss.
This dual approach enables the model to more accurately approximate the true solution. 
These observations are further illustrated in \figurename~\ref{fig:lorenz_forward_analysis}.
Note that evaluation losses are computed across the entire period, with the weighting coefficients $\lambda_i$ maintained at 1.0; $\text{OdePINN}_\text{orig}$ are not shown as the losses are at different scales and its approximation is similar to $\text{OdePINN}_\text{causal}$.
Fig \ref{fig:lorenz_forward_analysis}a demonstrates that the $\text{OdePINN}_\text{grad+causal}$ model achieves a significantly low final total loss of $1.6 \cdot 10^{-4}$, which is considerably less than that of the $\text{OdePINN}_\text{causal}$ at $2.2 \cdot 10^{-3}$. 
The remaining models display higher losses of 0.046 and 0.056, with the baseline model showing slightly better performance.
Comparison of errors across the models indicates that Gradient Balancing substantially enhances convergence towards the data loss term $\mathcal{L}_{data}$.
The data losses for $\text{OdePINN}_\text{grad}$ and $\text{OdePINN}_\text{grad+causal}$ are recorded at $1.4 \cdot 10^{-8}$ and $8.9 \cdot 10^{-9}$, respectively, which are significantly lower than the approximately $10^{-2}$ or $10^{-3}$ observed in the other models.
Despite achieving a significantly lower data loss, the total loss of $\text{OdePINN}_\text{grad}$ is $6.3 \cdot 10^{-2}$, slightly higher than the baseline model's $4.5 \cdot 10^{-2}$, due to a higher ODE loss.
This technique enables the neural network to better fit the initial condition but simultaneously makes it harder to satisfy the ODE loss, resulting in higher ODE and overall losses. Greater detail on the usage and analysis of this method is provided in \cite{Wang2020UnderstandingAM}.
Ultimately, the root mean squared error between the grad model and the true solution is 11.9, slightly better than the baseline's 12.18, though both remain far from a perfect solution.

However, the Gradient Balancing technique on its own does not address the issue of causal effects: Causal Training is essential in this case.
Gradient Balancing averages gradients from different times, leading to scenarios in Fig \ref{fig:lorenz_forward_analysis}b, where at later time $t=1.5$ or $t=2.0$, the model can quickly minimize the physics constraints to a negligible level of $10^{-5}$ at an early stage in the training.
Notably, within the domain $t \in [1.0, 2.0]$, the model converges to the null solution (as illustrated in \figurename~\ref{fig:lorenz_forward_u_pred}), which satisfies the ODE system.
However, at earlier times, such as $t=0$ or $t=0.5$, the model consistently struggles throughout the training duration.
We hypothesize that this premature convergence of the ODE constraints at later times traps the model in a local minimum, preventing it from satisfying the constraints at earlier times and thus resulting in an inaccurate solution approximation. 
To address this, we employ a Causal Training strategy that trains the physics-laws term using residual points drawn from a progressively expanding interval, starting from the 1,000th to the 200,000th step.
This growing interval strategy ensures that the model adheres to the system dynamics at earlier times before progressing to later times, thereby respecting the causal effects.
The domain in which the model complies with the differential equations broadens as the training advances, as depicted in Fig \ref{fig:lorenz_forward_analysis}c. 
Outside this domain, the model's behavior remains arbitrary.
Ultimately, the $\text{OdePINN}_\text{grad+causal}$ model effectively combines these techniques, effectively minimizing both data and physics losses to achieve a highly accurate solution approximation.

%
%

\subsection*{Forward Problem $T=20$}
For a more robust test, we also performed a tougher evaluation attempting to predict all values on a continuous scale between 0 and 20. We demonstrate domain decomposition by solving the Lorenz system over an extended domain, from $t=0$ to $T=20.0$. The initial condition is set as $(1, 1, 1)$, with constant, known parameters $\sigma = 10$, $\rho = 28$, and $\beta = \frac{8}{3}$.

\begin{figure}[!ht]
    \centering
    \includegraphics[width=1.0\linewidth]{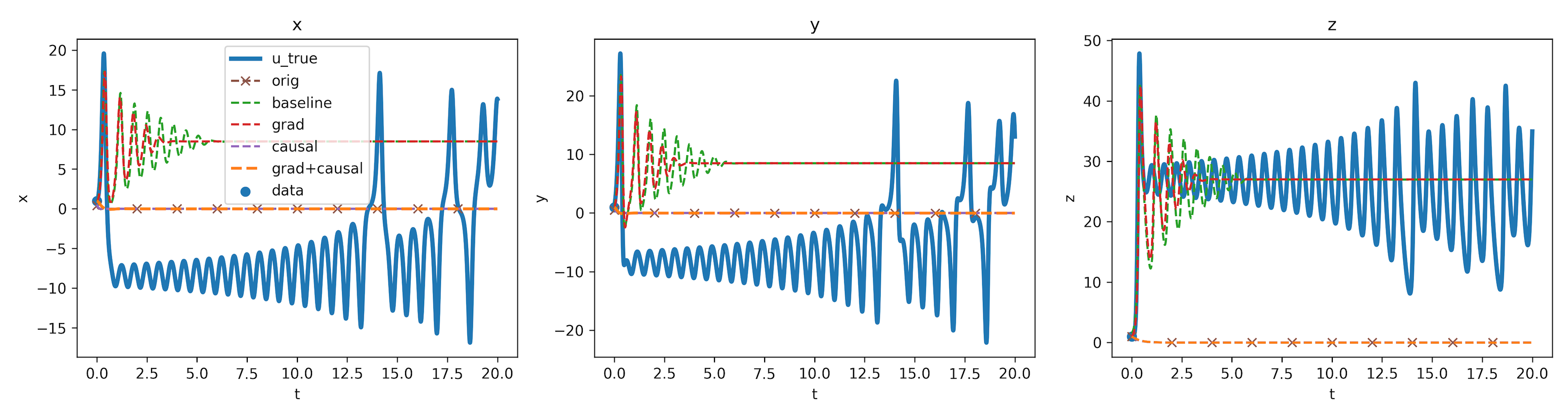}
    \caption{\textbf{Lorenz ODE system, forward problem, $U$ approximation of the system state from the first 5 models, using the first five models, excluding the Domain Decomposition model.}}
    \label{fig:lorenz_forward_T20}
\end{figure}

Fig~\ref{fig:lorenz_forward_T20} displays the the results when the domain is larger where none of the models achieve a close approximation to the reference solution.
The two models $\text{OdePINN}_\text{baseline}$ and $\text{OdePINN}_\text{grad}$ demonstrate some level of ability to learn the initial condition and marginally adhere to the ODE equations.
However, their errors grow sufficiently large by $t=0.75$, leading them to diverge significantly from the true solution, eventually converging to a trivial constant solution beyond $t=3$.
Other models, $\text{OdePINN}_\text{orig}$, $\text{OdePINN}_\text{causal}$ and $\text{OdePINN}_\text{grad + causal}$ exhibit a similar pattern as observed in the $T=2$ experiment, converging to the null solution.
The models' inaccuracies can be attributed to the increased domain size, which raises the solution's complexity. 
Therefore, it was necessary to apply Domain Decomposition to manage the larger domain size.

The domain is divided into 40 subdomains, each with a size of 0.6 and an overlap of 0.05 on both ends.
Each subdomain is independently modeled and trained by a distinct neural network. 
The predictions from the previous subdomain, uniformly distributed over 100 points within the overlapped region, serve as the data conditions for the subsequent subdomain.
The hyper-parameters remain identical across all subdomains.
The neural network $U$ is structured as an MLP with four hidden layers, each comprising 100 units and utilizing the GELU activation function.
The training process is limited to a maximum of 150,000 steps, starting with an initial phase of 5,000 steps dedicated to data fitting, followed by 100,000 steps focused on causal training and the remaining are for final tuning phase.
Additionally, the gradient balancing weights, $\lambda_i$, are updated every 100 steps, utilizing a smoothing factor of $\alpha=0.99$.

\begin{figure}[!ht]
    \centering
    \includegraphics[width=1.0\linewidth]{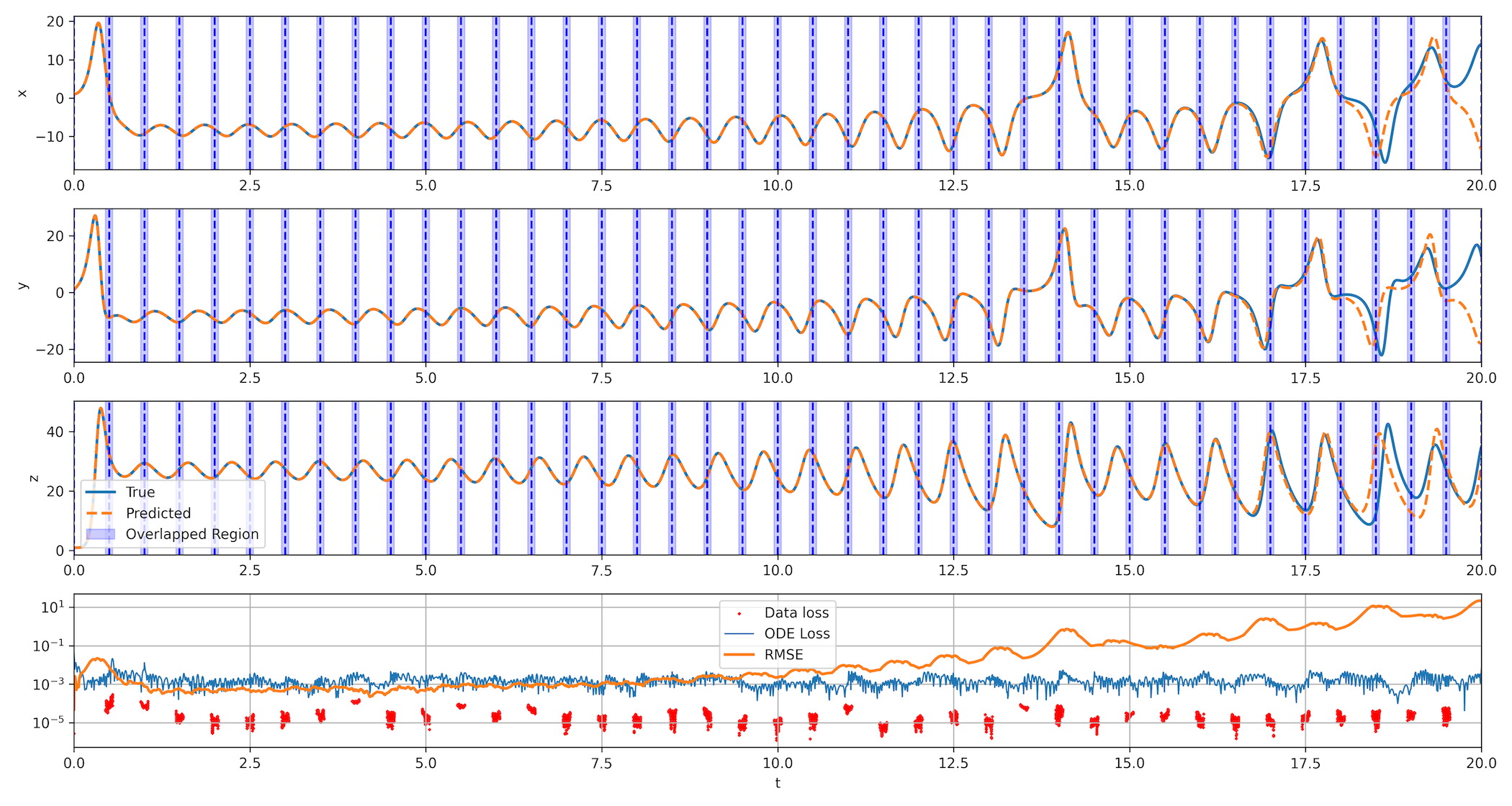}
    \caption{\textbf{Lorenz ODE system, forward problem, $U$ approximation of the system state, using the domain composition model.}}
    \label{fig:lorenz_forward_decomposition}
\end{figure}

\figurename~\ref{fig:lorenz_forward_decomposition} illustrates the solution approximated by the proposed framework, accompanied by plots of the training losses and Root Mean Squared Error (RMSE) relative to the ground truth data.
The model demonstrates a fairly good prediction of the system's evolution.
Initially, the RMSE is approximately $5 \cdot 10^{-4}$, but it exponentially increases as $t$ progresses, where the RMSE escalates to $10^{-3}$ at $t=8$, to $10^{-1}$ at $t=13$, and reaches an error magnitude of $10^1$ by the end of the period.
The final four subdomains, as depicted in \figurename~\ref{fig:lorenz_forward_decomposition} (12.5 to 20.0), show notable approximation errors.
This substantial error accumulation towards the end of the period is understandable given that the Lorenz system is highly sensitive to initial conditions where minor predictive inaccuracies can significantly deviate the future states.
As a result, initial training errors rapidly accumulate over time, culminating in an RMSE of up to 10 by the end of the time period.
The training data loss remains below $10^{-4}$ throughout, with many instances dropping to the $10^{-5}$ level.
Furthermore, the errors related to physical constraints are consistently maintained at the $10^{-3}$ level, indicating that the physical laws are satisfied with ODE loss approaching zero.
Notably, the losses peak in regions of the solution characterized by sharp changes, consequently resulting in a steep increase in RMSE.
On a positive front, the framework exhibits consistent performance across all subdomains. 
However, the clear accumulation of errors highlights a potential limitation of the technique, suggesting areas for further improvement.

\begin{figure}[!ht]
    \centering
    \includegraphics[width=1.0\linewidth]{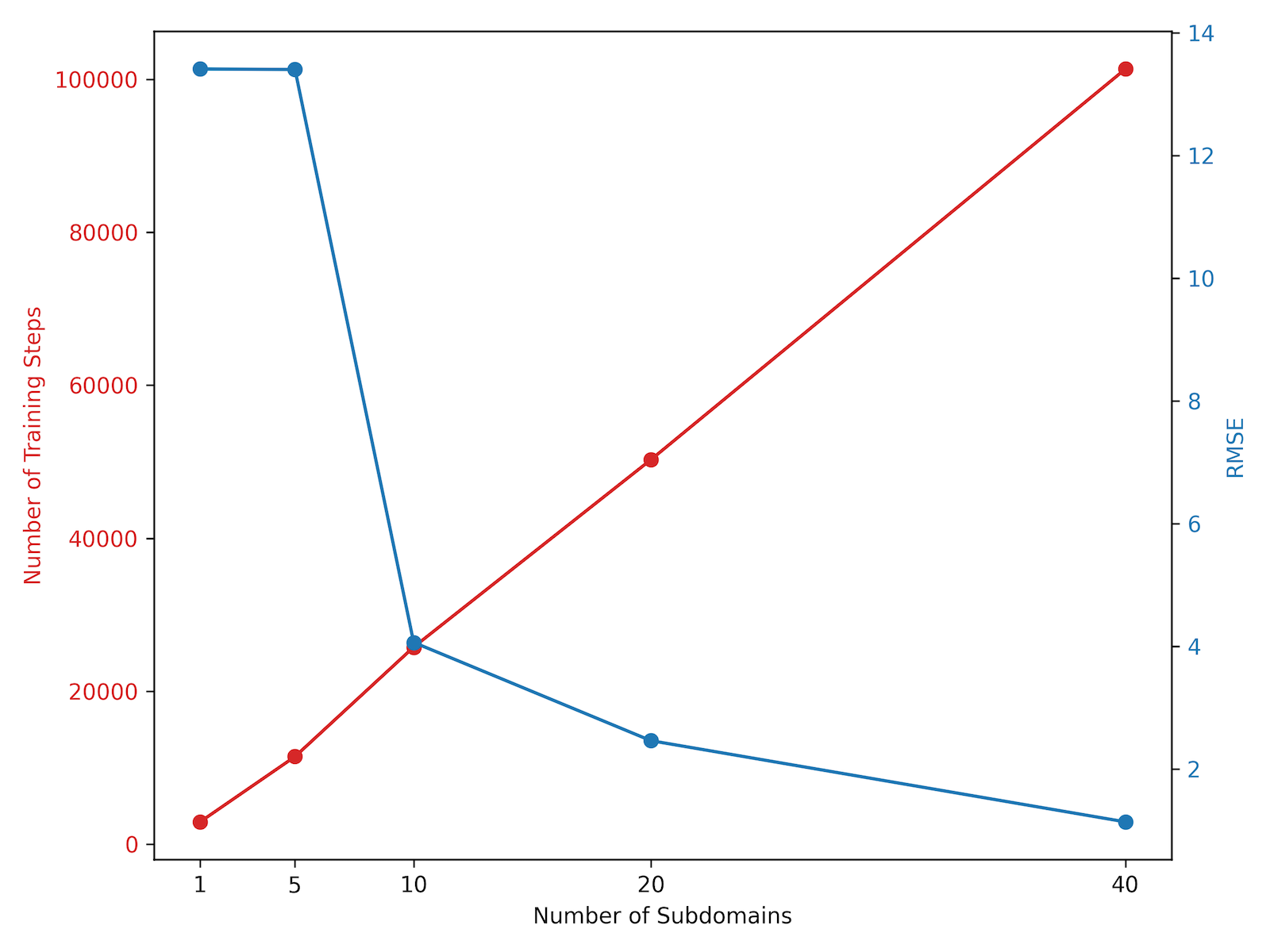}
    \caption{\textbf{Trade-off between the number of subdomains and accuracy.}}
    \label{fig:lorenz_forward_nsubdomains}
\end{figure}

\figurename~\ref{fig:lorenz_forward_nsubdomains} plots the relationship between the number of training steps and the RMSE across a range of subdomains.
The training steps increase linearly with the number of subdomains as each subdomain requires approximately 200,000 to 300,000 training steps, equivalent to around 30 minutes on the GPU NVIDIA GeForce RTX 4090 hardware used for all experiments.
The figure demonstrates a clear trade-off between the number of subdomains and model accuracy: as the number of subdomains rise, the RMSE decreases exponentially, but at the cost of longer training times.
When the number of subdomains is low, such as 1 or 5, the training domain remains too large, causing the model to converge to the null solution, yielding an RMSE of 13.4. 
With an increase in subdomains, starting from 10, the training domain becomes smaller, reducing the solution complexity within each subdomain.
This reduction leads to a RMSE decrease, reaching 4.1 with 10 subdomains and 1.13 with 40 subdomains.
These results indicate that cumulative errors can be mitigated by increasing the number of subdomains, allowing the model to focus on improving accuracy at finer levels of detail.

%
%
\subsection*{Inverse Problem}
\label{sec:lorenz_inverse}

\begin{figure}[!ht]
  \centering
  
    \includegraphics[width=1.0\linewidth]{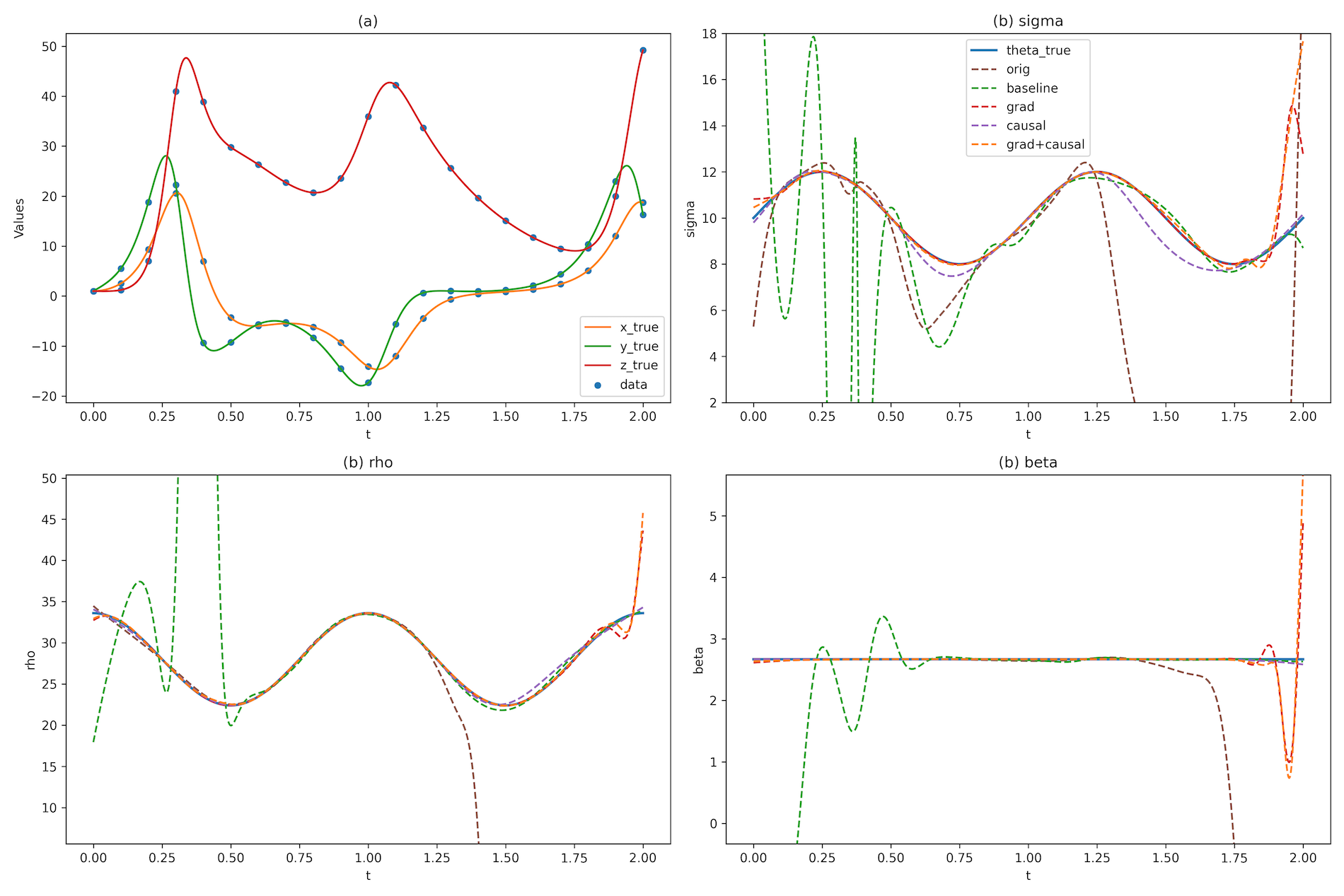}
  \caption{\textbf{OdePINN framework solves the Lorenz system inverse problem.}
  (a): Ground truth $u$ and data provided to PINN;
  (b): $\Theta$ approximation of the physics parameters $(\sigma, \rho, \beta)$.
  }
  \label{fig:lorenz_inverse}
\end{figure}

In this section, we illustrate the application of the proposed framework to an inverse problem scenario, using significantly more data. Here, the framework is employed to predict the physics parameters and simultaneously interpolating the system state from a limited number of observations of the system's state.
We conduct these experiments using the Lorenz system over the time domain $t \in [0, 2.0]$, with time-varying physical parameters defined as $\sigma = \frac{10}{2} \sin \left( 2 \pi t \right) + 10$, $\rho = \frac{28}{5} \sin \left( 2 \pi t + \frac{\pi}{2} \right) + 28$, and $\beta = \frac{8}{3}$.
These formulas are unknown to and are to be learnt by the framework.
The initial conditions are set to $(1, 1, 1)$, consistent with previous experiments.
The dataset for training comprises 21 simulated data points that are evenly distributed across the input domain, resulting in a dataset dimension of $(21, 3)$.
Fig \ref{fig:lorenz_inverse}a presents both the reference solution discussed earlier \cite{doi:10.1137/0904010} and the simulated data points.

The configuration of the framework is similar to previously described experiments. 
The main neural network $U$ retains its structure from earlier experiments, comprising an MLP with four hidden layers, each consisting of 100 units, with GELU activation. 
Additionally, three smaller, distinct neural networks $\Theta$ are implemented to model the three physical parameters of the dynamical system.
These networks are each equipped with four hidden layers containing 10 units, using GELU as the activation function for all hidden layers.
All networks incorporate the time variable $t$ as a singular input to capture the temporal dynamics of the parameters.
In terms of training enhancements, Gradient Balancing is used where the weights of the objective function, $\lambda_{i}$, are adjusted every $N=100$ steps, employing a smoothing factor of $\alpha=0.99$.
The Causal Training approach involves an initial data fitting phase comprising 10,000 steps, reflecting the increased data availability.
This is followed by progressive causal training and a final tuning phase, which are conducted over 200,000 and 100,000 steps, respectively.

\figurename~\ref{fig:lorenz_inverse}b displays the dynamical system's parameters as approximated by the neural network $\Theta$. 
While all models demonstrate an ability to capture the general trends of the parameters, they also exhibit inherent inaccuracies. 
Notably, the original and baseline models show substantial errors particularly in regions where the derivatives of $u$ with respect to time $t$ are remarkably high, as can be seen within the interval $[0, 0.5]$ or by $t=2.0$ in Fig \ref{fig:lorenz_inverse}a.
Among the models, $\text{OdePINN}_\text{grad}$ achieves the most accurate parameter approximations throughout most of the domain.
However, it encounters difficulties at the period's end, near $t=2.0$.
In this inverse setting, the system's parameters are not predefined and are subject to variation, leading to non-unique solutions for $u$ and $\theta$ in the absence of sufficient data.
Consequently, $\text{OdePINN}_\text{grad}$ tends to favor solutions with smaller magnitude derivatives, which can result in significant discrepancies when compared to the reference solution.
Conversely, $\text{OdePINN}_\text{causal}$ typically exhibits error accumulation towards higher values of $t$, with noticeable inaccuracies in estimating parameters such as $\sigma$ and $\rho$ beyond $t = 1.15$, as illustrated in \figurename~\ref{fig:lorenz_inverse}b.
This pattern of error propagation is a recurring observation throughout our study. 
The combined model, $\text{OdePINN}_\text{grad+causal}$, mirrors the performance of $\text{OdePINN}_\text{grad}$, maintaining reasonable accuracy until the domain's far end.
The approximation $U$ by the models can be found in  \nameref{sec:lorenz_inverse_theta_pred}.

Table \ref{tab:lorenz_inverse_errors} presents a detailed analysis of the errors associated with both the interpolation of $u$ and the estimation of parameters $\theta$, benchmarked against the ground truth data, errors calculated in metrics presented in \nameref{sec:metrics}.
The \textit{original} model registers highest errors, with RMSE values of 3.9564 for $u$ and 52.4141 for $\theta$, alongside MDAPE of 2.6219 and 24.5114, respectively.
Compared to the \textit{original} model,  the baseline model shows slight improvements but still incurs relatively high errors.
Conversely, $\text{OdePINN}_\text{grad}$ obtained significant improvements in accuracy, recording the lowest errors across all evaluated metrics for both $u$ and $\theta$. It achieves an RMSE of 0.2525 for $u$ and an impressively low 0.7195 for $\theta$. Additionally, it reports MAE values of 0.0472 and 0.1602, and MDAPE scores below 0.01\% for both variables.
The causal model records the best RMSE for $u$ at 0.1039, although its performance on other metrics does not reach the levels achieved by the grad model.  $\text{OdePINN}_\text{grad+causal}$, which integrates the approaches of the preceding two models, does not achieve top scores in any specific category but secures the second-best results, including impressive MDAPE values below 0.01\% and 0.06\%.
This final training stage of $\text{OdePINN}_\text{grad+causal}$ mirrors the training process of the entire $\text{OdePINN}_\text{grad+causal}$ model, albeit with an improved initialization and a reduced number of training steps.

\begin{table}[!ht]
\begin{adjustwidth}{-2.00in}{0in}
\centering
\caption{Approximation errors in the inverse problem with Lorenz system where smallest error values are best.}
\label{tab:lorenz_inverse_errors}
\begin{tabular}{llrrrr|rrrr}
\toprule
Model       & Metric                 & \multicolumn{1}{c}{$x$} & \multicolumn{1}{c}{$y$} & \multicolumn{1}{c}{$z$} & \multicolumn{1}{c|}{$u$}         & \multicolumn{1}{c}{$\sigma$} & \multicolumn{1}{c}{$\rho$} & \multicolumn{1}{c}{$\beta$} & \multicolumn{1}{c}{$\theta$}     \\ \midrule
orig & \multirow{5}{*}{RMSE} & 4.4453 & 4.9325 & 1.6939 & 3.9564 & 5.4294 & 90.5381 & 3.8850 & 52.4141 \\
baseline    &   & 3.4995                & 2.8182                & 0.6591                & 2.6219                & 8.4507                    & 41.5828                 & 1.3683                   & 24.5114                   \\
grad        &                        & 0.1011                & 0.3556                & 0.2336                & {\ul 0.2525}          & 0.8727                    & 0.8498                  & 0.2629                   & \textbf{0.7195}           \\
causal      &                        & 0.1363                & 0.0949                & 0.0692                & \textbf{0.1039}       & 1.3853                    & 2.4592                  & 0.0740                   & 1.6302                    \\
grad+causal &                        & 0.0862                & 0.3760                & 0.2659                & 0.2705                & 1.2593                    & 0.8922                  & 0.3316                   & {\ul 0.9114}              \\ \midrule
orig & \multirow{5}{*}{MAE} & 1.7002 & 1.0028 & 0.3667 & 1.0232 & 3.6059 & 47.0012 & 1.4000 & 17.3357 \\
baseline    &    & 0.9579                & 0.6687                & 0.2430                & 0.6232                & 3.1751                    & 10.9855                 & 0.5066                   & 4.8891                    \\
grad        &                        & 0.0207                & 0.0759                & 0.0449                & \textbf{0.0472}       & 0.2116                    & 0.2078                  & 0.0611                   & \textbf{0.1602}           \\
causal      &                        & 0.0844                & 0.0654                & 0.0452                & 0.0650                & 0.7099                    & 1.3002                  & 0.0472                   & 0.6858                    \\
grad+causal &                        & 0.0192                & 0.0904                & 0.0543                & {\ul 0.0546}          & 0.2845                    & 0.2590                  & 0.0827                   & {\ul 0.2087}              \\ \midrule
orig & \multirow{5}{*}{MDAPE} & 0.0063 & 0.0064 & 0.0015 & 0.0025 & 0.1271 & 0.0127 & 0.0058 & 0.0200 \\
baseline    &  & 0.0019                & 0.0029                & 0.0016                & 0.0019                & 0.0382                    & 0.0114                  & 0.0086                   & 0.0162                    \\
grad        &                        & 0.0001                & 0.0002                & 0.0001                & \textbf{0.0001}       & 0.0005                    & 0.0005                  & 0.0003                   & \textbf{0.0004}           \\
causal      &                        & 0.0004                & 0.0012                & 0.0016                & 0.0013                & 0.0175                    & 0.0044                  & 0.0074                   & 0.0090                    \\
grad+causal &                        & 0.0000                & 0.0001                & 0.0001                & \textbf{0.0001}       & 0.0006                    & 0.0003                  & 0.0007                   & {\ul 0.0006}              \\ \midrule
orig & \multirow{5}{*}{nRMSE} & 0.2562 & 0.2450 & 0.0703 & 0.2086 & 2.7147 & 16.1675 & 3.8850 & 9.7271 \\
baseline &  & 0.2017 & 0.1400 & 0.0274 & 0.1426 & 4.2253 & 7.4255 & 1.3683 & 4.9955 \\
grad &  & 0.0058 & 0.0177 & 0.0097 & {\ul 0.0121} & 0.4364 & 0.1517 & 0.2629 & {\ul 0.3069} \\
causal &  & 0.0021 & 0.0025 & 0.0008 & \textbf{0.0019} & 0.2210 & 0.0608 & 0.0156 & \textbf{0.1326} \\
grad+causal &  & 0.0079 & 0.0162 & 0.0140 & 0.0132 & 0.5519 & 0.1572 & 0.3124 & 0.3772 \\ \midrule
\end{tabular}
\begin{flushleft} \textbf{Bold text} highlight the best performing models with respect to a specific metric and \uline{Underlined numbers} represent the second best performing model.
\end{flushleft}
\end{adjustwidth}
\end{table}


With an increase in data availability, the preliminary two phases of causal training contribute little to overall improvement in performance.
This is mainly because a dataset of sufficient volume can establish a robust initialization, diminishing the relative advantage of the first two phases.
On the other hand, $\text{OdePINN}_\text{grad+causal}$, with its shorter training duration, does not attain performance levels comparable to those of the Gradient-Balancing model.

%
%
%
%
\section*{Case Study-Based Validation}
\label{sec:mosquito}
In addition to a theoretical validation, our PINN framework is also validated in the practical environment of dynamical modelling of the mosquito population. Here, our approach is adopted on the ODE-based model of mosquito population dynamics proposed in \cite{CAILLY20127}. 
The model divides the mosquito life cycle into 10 stages: Egg ($E$), Larva ($L$), Pupa ($P$), Emerging Adults ($A_{em}$), Nulliparous Bloodseeking Adults ($A_{b1}$), Nulliparous Gestating Adults ($A_{g1}$), Nulliparous Ovipositing Adults ($A_{o1}$), Parous Bloodseeking Adults ($A_{b2}$), Parous Gestating Adults ($A_{g2}$) and Parous Ovipositing Adults ($A_{o2}$).
The 10 stages are related via ordinary equations as show in Eq~\eqref{eq:mosquito_ode_system}, with parameters explained in \nameref{sec:mosquito_ode_parameters}.
A diagram illustrating the stages and transitions is available in \nameref{sec:modscheme}.
System parameters are modelled as a function depending on temperature with data acquired for \textit{Culex pipiens} spieces from \cite{petric2020}.

\begin{align}
    \left\{
    \begin{array}{ll}
        \frac{dE}{dt} &= \gamma_{Ao} ( \beta_1 A_{o1} + \beta_2 A_{o2})  - (\mu_E + f_E) E\\
        \frac{dL}{dt} &= f_E E - \left( m_L \left( 1 + \frac{L}{\kappa_L} \right) + f_L \right) L \\
        \frac{dP}{dt} &= f_L L - (m_P + f_P) P \\
        \frac{d A_{em}}{dt} &= f_P \sigma e^{-\mu_{em} \left( 1 + \frac{1}{\kappa_P} \right)} P - (m_A + \gamma_{Aem}) A_{em} \\
        \frac{dA_{b1}}{dt} &= \gamma_{em} A_{em} - (m_A + \mu_r + \gamma_{Ab}) A_{b1} \\
        \frac{d A_{g1}}{dt} &= \gamma_{Ab} A_{b1} - (m_A + f_{Ag}) A_{g1} \\
        \frac{d A_{o1}}{dt} &= f_{Ag} A_{g1} - (m_A + \mu_r + \gamma_{Ao}) A_{o1} \\
        \frac{d A_{b2}}{dt} &= \gamma_{Ao} (A_{o1} + A_{o2}) - (m_A + \mu_r + \gamma_{Ab}) A_{b2} \\
        \frac{d A_{g2}}{dt} &= \gamma_{Ab} A_{b2} - (m_A + f_{Ag}) A_{g2} \\
        \frac{d A_{o2}}{dt} &= f_{Ag} A_{g2} - (m_A + \mu_r + \gamma_{Ao}) A_{o2}
    \end{array}
    \right.
    \label{eq:mosquito_ode_system}
\end{align}

This part of the evaluation comprised two challenges: first, solving the mosquito population dynamics with an initial condition; and second, determining mortality and growth rates over time from available data. We conducted tests under a varying temperature condition, where the temperature changes according to a sine function $\tau = 10 \sin \left( 2 \pi \frac{t}{365} \right) + 10$ with time, $t$, measured in days. 

\subsection*{Forward Problem}

For the forward problem, the goal was to solve the mosquito population dynamics using only a single data point at the beginning, known as the initial condition. 
Specifically, we employ a numerical method \cite{doi:10.1137/0904010} for solving ordinary differential equations (ODEs), available in Python Scipy library \cite{2020SciPy-NMeth}
, to simulate data over three years ($t$ ranging from $0$ to $1096$ days). 
We use the data point at $t = 730$ (two years into the simulation) as the starting point and trained a PINN for the period from $730$ to $1,096$ days. 
The simulated data within this period serves as the ground truth solution against which we evaluate the model’s performance. The advanced techniques presented in the section \nameref{sec:methodology}, including ODE normalization, domain decomposition, gradient balancing and causal training, were implemented.

The time period was separated into 12 subdomains where for each, training was carried out for 10,000 steps to fit the data, followed by 100,000 steps focusing on causal training. 
Early stopping was used after 100 evaluations without improvement in the final phase. 
The lower and upper bounds for each domain are acquired from the ground truth data.
The solution neural network $U$ was configured as an MLP with 4 hidden layers with 100 units each GELU as activation function for every hidden layer.
The subdomains are trained sequentially, with the predictions in the overlapped domain of previous subdomain models used as the initial data condition for the next domain.

\begin{figure}[!ht]
    \centering
    \includegraphics[width=1.0\linewidth]{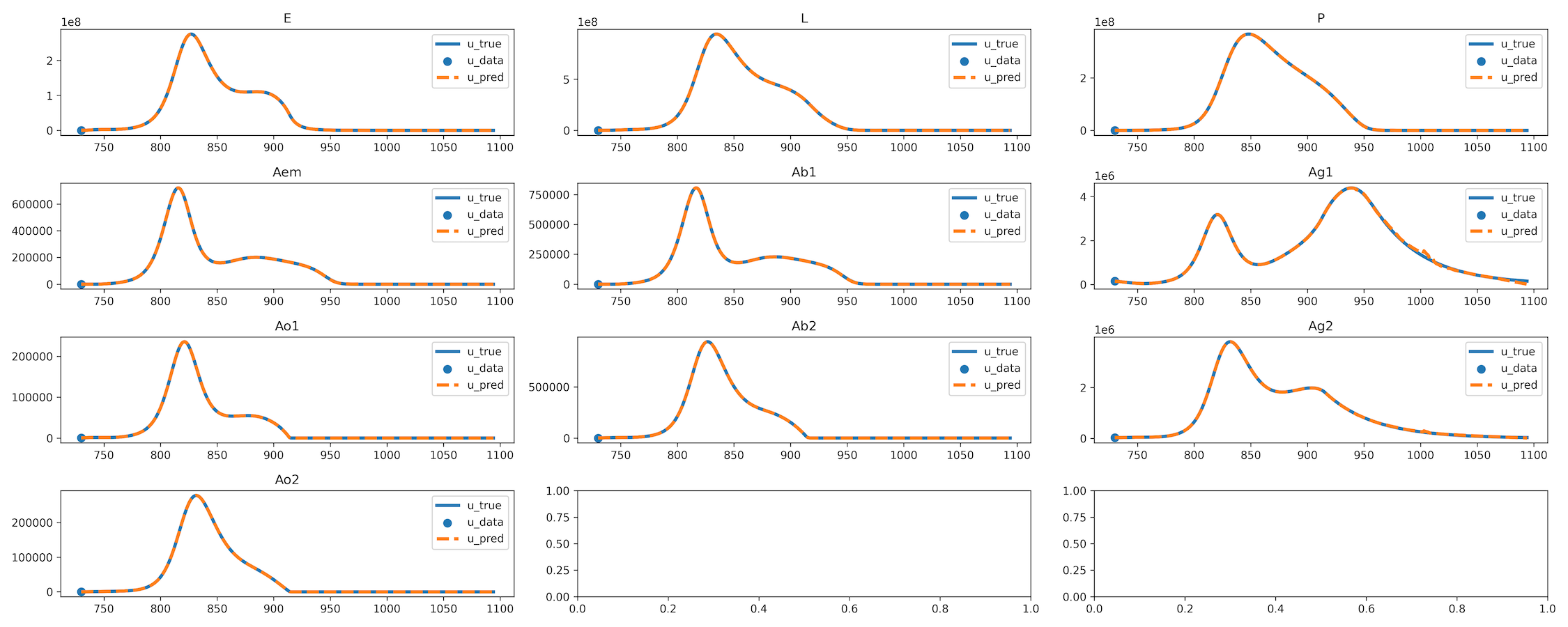}
    \caption{\textbf{Mosquito ODE system, forward problem, $U$ approximation of the solution.}}
    \label{fig:forward_t=sin}
\end{figure}

The results for solving the mosquito population dynamics as a forward problem are shown in \figurename~\ref{fig:forward_t=sin}, with error measurements provided in Table \ref{tab:mosquito_forward_t=sin_errors}.
The trained neural network shows impressive performance in extrapolating and providing a fairly accurate approximation of the solution in \figurename~\ref{fig:forward_t=sin}. 
However, towards the end of the time period, accumulated error can be seen in Ag1 and Ag2 stages of mosquitoes development.

\begin{table}[!ht]
\centering
\caption{Approximation Errors for Mosquito ODE Solution.}
\label{tab:mosquito_forward_t=sin_errors}
\begin{tabular}{lrrrr}
\toprule
Stage & RMSE & MAE & MDAPE & nRMSE\\
\midrule
E & 56107.947877 & 28619.218663 & 0.001900 & 0.000407 \\
L & 144952.348476 & 82367.593703 & 0.000759 & 0.000308 \\
P & 243295.690709 & 123936.043785 & 0.002170 & 0.001325 \\
Aem & 398.308537 & 175.715612 & 0.001821 & 0.001106 \\
Ab1 & 342.155798 & 161.987210 & 0.001782 & 0.000850 \\
Ag1 & 42095.722155 & 20403.707941 & 0.002158 & 0.019350 \\
Ao1 & 34.564301 & 16.196719 & 0.003921 & 0.000293 \\
Ab2 & 102.456176 & 50.194130 & 0.002831 & 0.000217 \\
Ag2 & 7853.480625 & 4466.414135 & 0.001567 & 0.004155 \\
Ao2 & 19.229490 & 9.978838 & 0.002394 & 0.000138 \\
Overall & 92296.312508 & 26020.705074 & 0.001918 & 0.006291 \\
\bottomrule
\end{tabular}
\end{table}

In Table \ref{tab:mosquito_forward_t=sin_errors}, the overall RMSE is quite high at 92,296 (organisms). 
Errors vary significantly across different life stages of mosquitoes, ranging from as low as 19 in $Ao2$ stage to as high as 243,295 at the $P$ stage, most likely due to the different scales of these life stages
The MAE shows a similar range of variation, from 10 ($Ao2$ stage) to 123,936 ($P$ stage), with an average of 26,021.
Despite these high values for RMSE and MAE, the Median Absolute Percentage Error (MDAPE), is relatively low at an average of 0.19 \%, which indicates a good performance in general from a machine learning perspective. 
MDAPE tends to be lower for stages with larger scales, suggesting that stages with larger scales are more tolerant of minor errors.
For instance, the smallest MDAPE is 0.0759\% for the $L$ stage, which has the second highest values in both RMSE and MAE. Conversely, the highest MDAPE, at 0.3921\%, occur in the $Ao1$ stage, which has relatively lower RMSE and MAE.

\subsection*{Inverse Problem}
In solving inverse problems, the goal was to predict 10 specific parameters using available data.
Three parameters ($\gamma_{Aem}$, $\gamma_{Ab}$ and $\gamma_{Ao}$) are treated as constants and are represented by learnable parameters that remain constant over time.
The remaining seven parameters ($f_E$, $f_P$, $f_L$, $f_{Ag}$, $m_L$, $m_P$ and $m_A$) vary over time with neural networks using time as input to estimate these parameters.
This experiment was conducted under the same conditions and temperatures as previous experiments but here, it was necessary to generate simulation data over a three-year period, from day 0 to day 1096, using the numerical method in \cite{doi:10.1137/0904010}.
For the data condition loss, we used daily data from $t=730$ to $t=1096$, totaling 367 days times 10 data points.
Both the solution $u$ and the ODE parameters $\theta$ were obtained from the simulation data as ground truth.

With this setup, the high data volumes allow for the use of simpler techniques.
Due to significant differences in the number of instances across various stages, ODE normalization and gradient balancing are applied.
Techniques such as domain decomposition and causal training were not used (set the number of subdomains to 1 and the number for the first two phases to 0), as the data condition with sufficient data provides a solid base for the framework's convergence.
The boundaries for the stages and ODE parameters were determined based on the data from the simulations.
For the neural network solution $U$, an MLP with 4 hidden layers, each containing 100 units, and utilizing the GELU activation function, was created. 
The constant parameters are represented by individual trainable weights.
Time-dependent parameters, on the other hand, are each represented by a smaller MLP, consisting of four hidden layers with 10 units per layer and also using the GELU activation function.

\begin{figure}[!ht]
    \centering
    \includegraphics[width=1.0\linewidth]{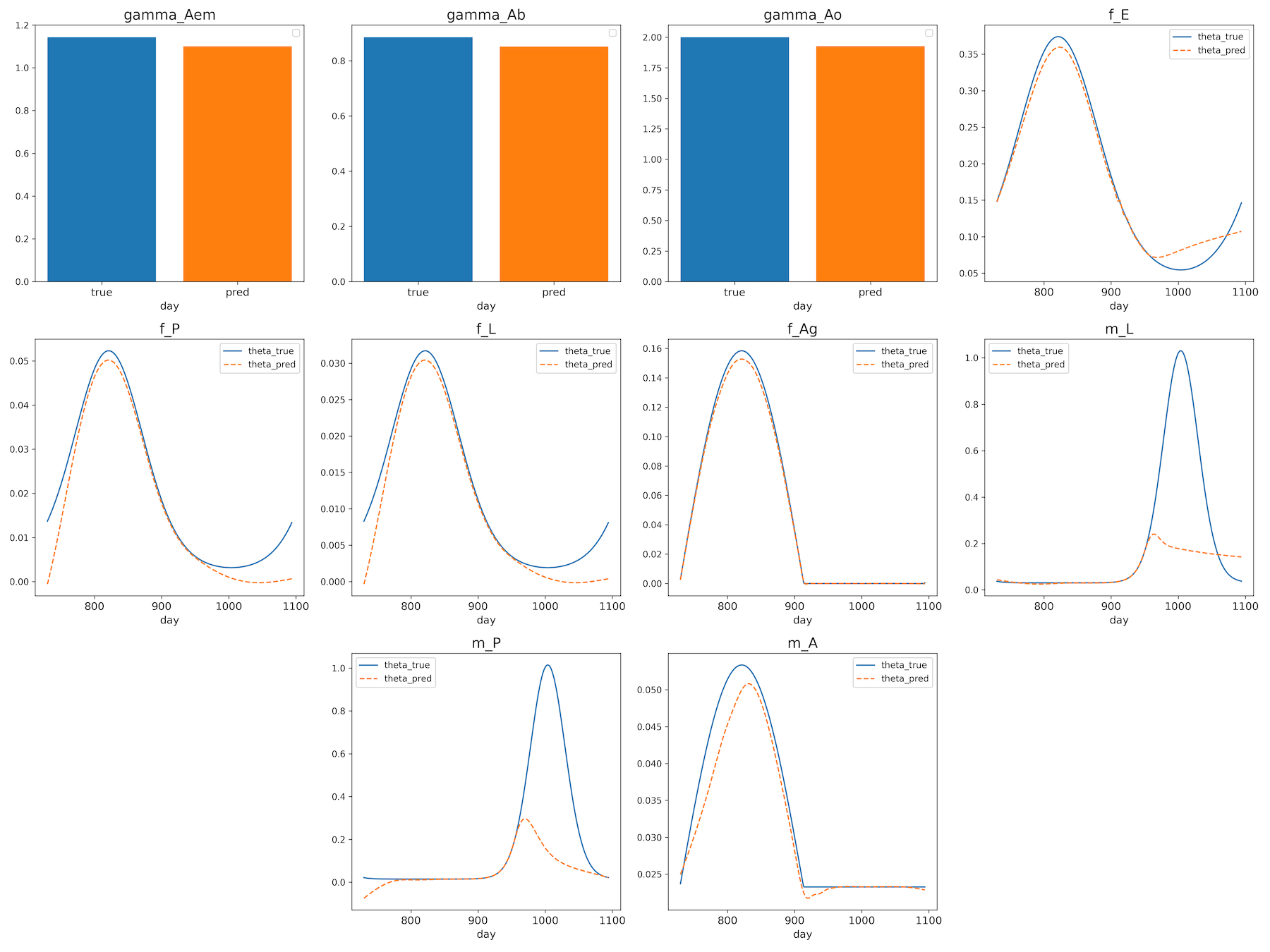}
    \caption{\textbf{Mosquito ODE system, inverse problem, $\Theta$ approximation of the system's parameters.}}
    \label{fig:inverse_t=sin}
\end{figure}

\figurename~\ref{fig:inverse_t=sin} shows the system parameters learnt from the PINN framework. 
The plots generally illustrate accurate parameter approximations although there are noticeable inaccuracies, particularly for parameters $m_L$ and $m_P$ within the domain $[943, 1064]$. 
The error metrics detailed in Table \ref{tab:mosquito_inverse_t=sin_errors}, reveal an average RMSE of 0.132615, with $m_L$ and $m_P$ contributing the highest errors at 0.286959 and 0.291548, respectively. 
The lowest RMSE values are for $f_{Ag}$ and $f_{L}$ at 0.002769 and 0.002816, respectively.
The MAE closely follows the RMSE trends, averaging at 0.044757. The highest MAE values correspond to $m_L$ and $m_P$ (0.137099 and 0.139703), and the lowest to $f_{Ag}$ and $f_{L}$ (0.001982 and 0.001784).
The MDAPE averages at 5.9\%, with $m_A$ showing the lowest error at 3.3566\% and the three constants ($\gamma_{Aem}$, $\gamma_{Ab}$, and $\gamma_{Ao}$) ranging between 3.62\% and 3.87\%.
The errors for $m_L$ and $m_P$ are notably higher at 12.59\% and 25.91\%, respectively.

\begin{table}[!ht]
\centering
\caption{Mosquito ODE System's Parameter Approximation Errors}
\label{tab:mosquito_inverse_t=sin_errors}
\begin{tabular}{lrrrr}
\toprule
Parameter & RMSE & MAE & MDAPE & nRMSE \\
\midrule
$\gamma_{Aem}$ & 0.042639 & 0.042639 & 0.037304 & 0.042639 \\
$\gamma_{Ab}$ & 0.034216 & 0.034216 & 0.038662 & 0.034216 \\
$\gamma_{Ao}$ & 0.072417 & 0.072417 & 0.036208 & 0.072417 \\
$f_E$ & 0.015570 & 0.012633 & 0.039964 & 0.015570 \\
$f_P$ & 0.004646 & 0.003270 & 0.078140 & 0.004646 \\
$f_L$ & 0.002816 & 0.001982 & 0.078126 & 0.002816 \\
$f_{Ag}$ & 0.002769 & 0.001784 & 0.094168 & 0.002769 \\
$m_L$ & 0.286959 & 0.137099 & 0.125888 & 0.286959 \\
$m_P$ & 0.291548 & 0.139703 & 0.259069 & 0.291548 \\
$m_A$ & 0.002848 & 0.001828 & 0.033566 & 0.002848 \\
Overall & 0.132615 & 0.044757 & 0.059045 & 0.132615 \\
\bottomrule
\end{tabular}
\end{table}

We interpret these results as follows: in colder temperatures, within the day range of 943 to 1064, where calculated air temperature is below $5.0 ^{\circ} C$, the mosquito population decreases rapidly to zero (with the exception of $Ag1$ and $Ag2$) as can be seen in \figurename~\ref{fig:forward_t=sin}.
This situation makes it challenging to gather useful information, negatively affecting the ability to determine several system parameters, especially $m_L$ and $m_P$ as these parameters are primarily derived from equations related to $\frac{dL}{dt}$ and $\frac{dP}{dt}$. 
Conversely, parameters $\gamma_{Aem}$, $\gamma_{Ab}$ and $\gamma_{Ao}$ remain constant over time, which makes them simpler to accurately estimate, as seen in their relatively low MDAPE.
Meanwhile, parameters like $m_A$, $f_{Ag}$ which are involved in several differential equations, tend to receive more consistent and stable information and gradients. 
In fact, the target parameters are not identifiable across the entire input domain due to the system setup (see \nameref{sec:identifiability}), which can result in inaccuracies in the approximations made by PINNs.

%
%
%
%

\section*{Discussion and Future Work}
\label{sec:discussion}

In this work, we introduced a hybrid framework based on Physics informed Neural Networks that integrates physical laws into data-driven machine learning models. 
This particular framework is designed to manage systems of ordinary differential equations, validated using a case study modelling mosquito population dynamics. 
The approach employs a multi-task learning strategy, incorporating multiple components in the objective function: one for data fitting, several for weakly enforcing physical knowledge, with separate functions for the remaining conditions.
The framework includes several advanced techniques such as domain decomposition, ODE normalization, gradient balancing, and causal training. 
We evaluated the effectiveness of our approach through an ablation study using the Lorenz system before introducing the problem of a model for the dynamics of mosquito populations.

The application of PINNs to mosquito population dynamics represents a significant advancement in ecological modelling for a number of reasons. Traditional data-driven models often require extensive data and are limited to capturing the nonlinear, multi-scale behaviours seen in real-world mosquito populations. However, PINNs integrate domain-specific knowledge directly into the learning process by enforcing ODE constraints. This is crucial for mosquito population dynamics because it ensures that the model respects biological principles even with limited data. 

Our findings demonstrate that ODE Normalization and Gradient Balancing played an important role in stabilizing the training process.
These techniques ensured that no individual component of the loss function disproportionately influenced the optimization, thus preventing premature convergence to suboptimal solutions.
Causal Training preserves the temporal causality inherent in the dynamical system, critical for achieving accurate model predictions, especially in scenarios where extrapolation beyond the scope of the training data is necessary.
Domain Decomposition effectively manages significantly large input domains, particularly in forward problem scenarios.
The results also confirm the framework's efficacy in modelling mosquito population dynamics, highlighting its potential for application to other ODE-based ecological models, which are commonly applied not only to mosquito populations but also to other vector species, such as ticks \cite{wu2020structured}, also agricultural and forest pests, including \textit{Drosophila suzukii} \cite{rossini2021general}, \textit{Batrocera oleae} \cite{rossini2022physiologically} and bark beetles \cite{kvrivan2016dynamical}.

Our research using PINNs revealed certain limitations.
In the inverse problem setup, the PINN tended to favor solutions with smaller gradients that still met the data and ODE constraints.
Furthermore, the capability of the framework to extrapolate in inverse problems remains uncertain; with lack of appropriate volume of data, the networks may converge to arbitrary solutions.
This aspect underscores the need for incorporating additional regularization techniques or constraints in order to guide the network towards more plausible solutions.
In the current implementation, the models consider only time as an input, without accounting for external factors. 
This restriction significantly limits predictive performance in real-world applications, as the coordination inputs and the encoded physical laws may not fully capture the underlying mechanisms of the processes.
Allowing for external variables would not only enhance performance but also add flexibility to the framework.
Although the framework has shown promising results on test systems, its scalability and ability to generalize to other types of dynamical systems have not been thoroughly evaluated. Our current focus is on the development of a new PINN framework that encodes complex, learned interactions of air temperature, precipitation and relative humidity, to improve and demonstrate the effectiveness of this approach in addressing inverse problems, such as inferring mosquito development and mortality rates.
Accurate mosquito population dynamics modelling is essential for predicting the outbreaks of arbovirus diseases. By improving predictive accuracy, PINNs can provide more reliable insights into population trends and help design targeted control measures (e.g., optimal times for pesticide application). This leads to more effective, data-informed vector control strategies that can be adapted across different geographic and climatic contexts.


\section*{Supporting information}

\paragraph*{S1 Appendix.}
\label{sec:metrics}
\textbf{Error metrics.}

Let $y_i,i=1\dots,M$ be $M$ reference values and $\hat{y}_i,i=1\dots,M$ are the corresponding predictions. Let $\mathfrak{L}$ and $\mathfrak{U}$ be the pre-defined lower and upper bounds for the values. The error metrics used in this paper, including Root Mean Squared Error (RMSE), Mean Absolute Error (MAE), Median Absolute Percenrage Error (MDAPE) and Root Mean Squared normalized Error (nRMSE) are defined as followed
\begin{align}
    RMSE &= \sqrt{\frac{1}{M} \sum_i^M \left( y_i - \hat{y}_i \right) ^ 2} \\
    MAE &= \frac{1}{M} \sum_i^M \left| y_i - \hat{y}_i \right| \\
    MDAPE &= \text{median}_i \left( \frac{\left| y_i - \hat{y}_i \right|}{\left| y_i\right|} \right)  \\
    nRMSE &= \frac{RMSE}{\mathfrak{U} - \mathfrak{L}}
\end{align}

\begin{figure}[h]
  \centering
    \includegraphics[width=\linewidth]{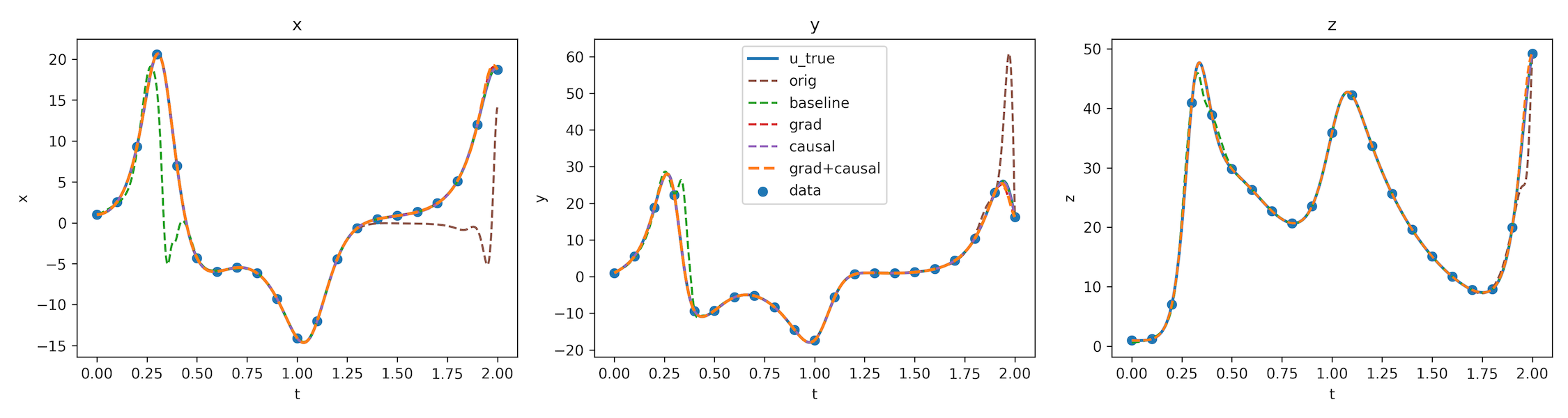}\\
  \caption{OdePINN framework solves inverse problem with Lorenz system, $U$ approximation of the solution.}
    \label{fig:s1_fig}
\end{figure}

\paragraph*{S1 Fig.}
\label{sec:lorenz_inverse_theta_pred} 
\textbf{OdePINN framework solves inverse problem with Lorenz system, $U$ approximation of the solution (\figurename~\ref{fig:s1_fig}).}

\paragraph*{S1 Table.} 

\label{sec:mosquito_ode_parameters}
\textbf{ODE Parameters (Table \ref{tab:mosquito_ode_parameters}).}

\begin{figure}[H]
  \centering
    \includegraphics[width=\linewidth]{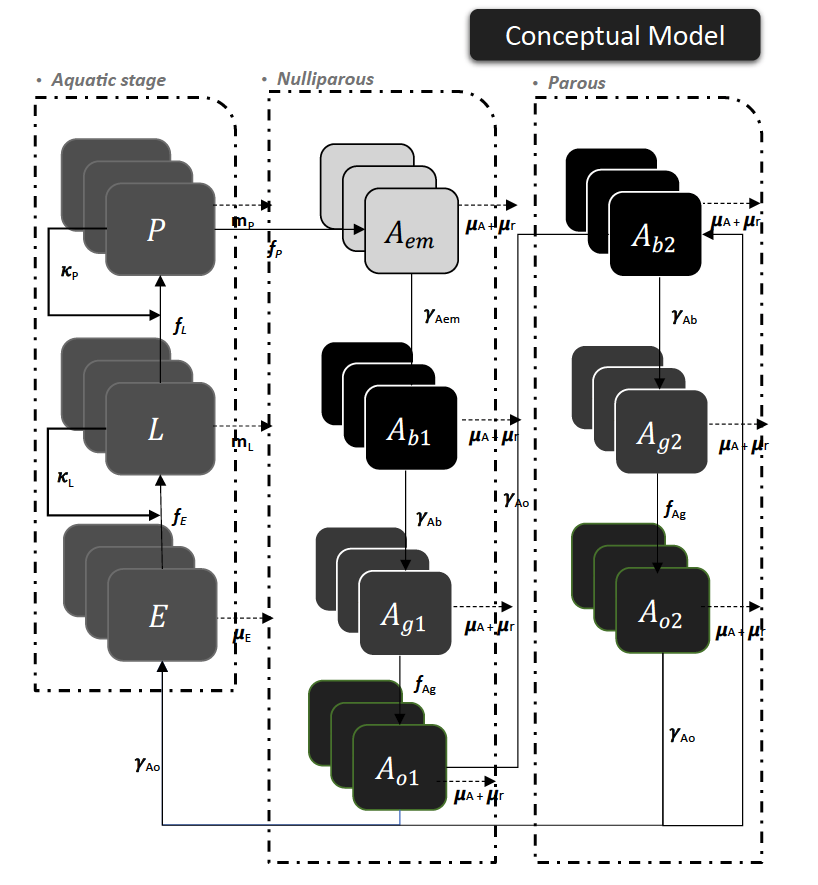}\\
  \caption{Model scheme showing the 10 develpmental stages within the mosquito life cycle: Egg ($E$), Larva ($L$), Pupa ($P$), Emerging Adults ($A_{em}$), Nulliparous Bloodseeking Adults ($A_{b1}$), Nulliparous Gestating Adults ($A_{g1}$), Nulliparous Ovipositing Adults ($A_{o1}$), Parous Bloodseeking Adults ($A_{b2}$), Parous Gestating Adults ($A_{g2}$) and Parous Ovipositing Adults ($A_{o2}$)  (source: \cite{petric2020})}
    \label{fig:modscheme}
\end{figure}

\paragraph*{S2 Fig.}
\label{sec:modscheme}
\textbf{Model scheme showing the 10 develpmental stages within the mosquito life cycle (\figurename~\ref{fig:modscheme})}: Egg ($E$), Larva ($L$), Pupa ($P$), Emerging Adults ($A_{em}$), Nulliparous Bloodseeking Adults ($A_{b1}$), Nulliparous Gestating Adults ($A_{g1}$), Nulliparous Ovipositing Adults ($A_{o1}$), Parous Bloodseeking Adults ($A_{b2}$), Parous Gestating Adults ($A_{g2}$) and Parous Ovipositing Adults ($A_{o2}$)  (source: \cite{petric2020}).

\begin{figure}[H]
  \centering
    \includegraphics[width=\textwidth]{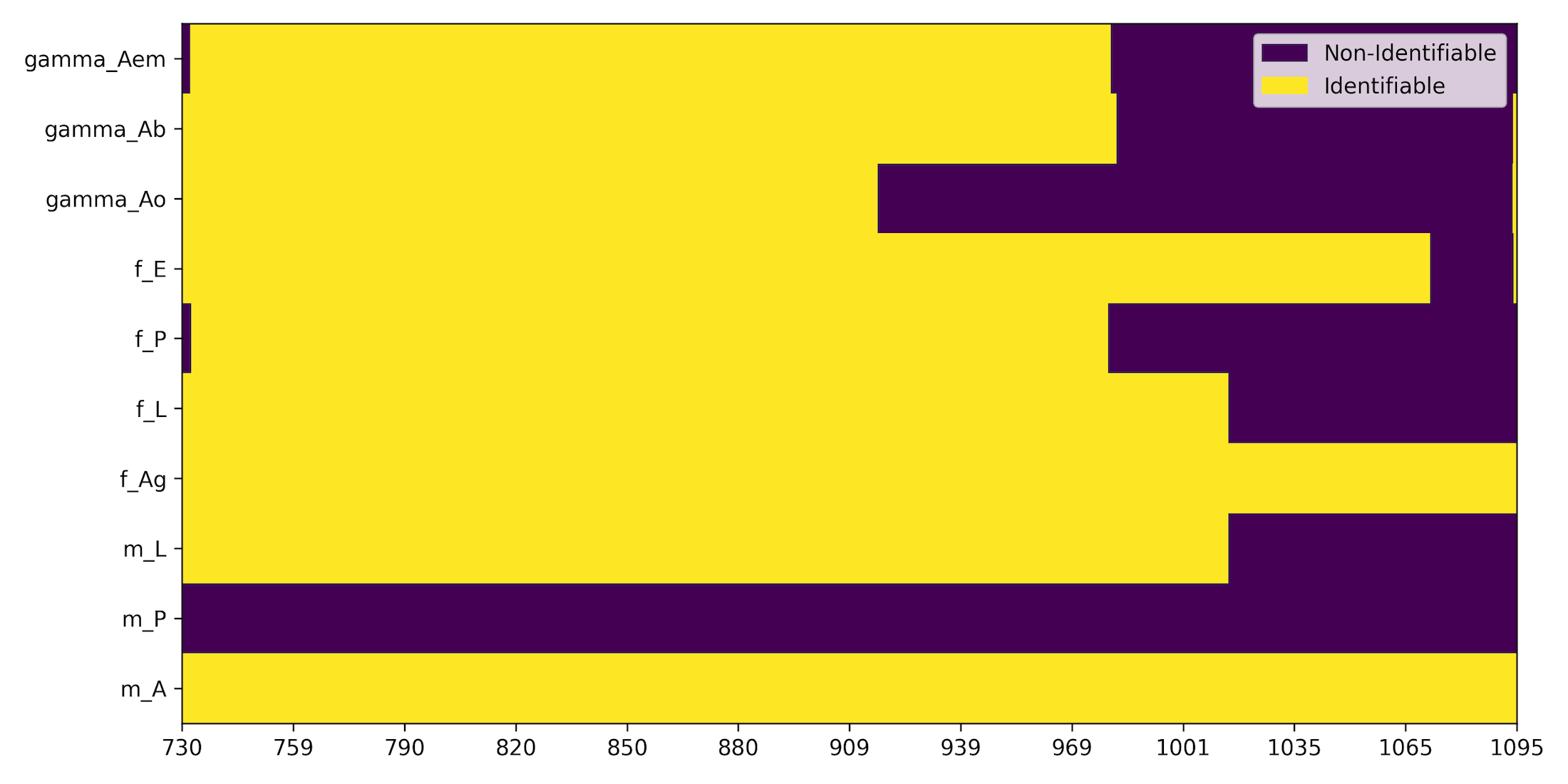}\\
  \caption{\textbf{Identifiability of parameters over time in the Mosquito inverse problem, the sine-shape temperature.}
This is achieved by expressing the system as a system of linear equations, the parameters as unknown variables, and analyzing the Reduced row-echelon form of the coefficient matrix, performed separately for each time $t$.
In the figure, identifiable parameters are defined  as free parameters which can get arbitrary values.
The values are rounded to 6-digit precision, aligning with the PINN's level of precision after training.
The figure explains the PINN's inaccuracies shown in \figurename~\ref{fig:inverse_t=sin}.
  }
    \label{fig:identifiability}
\end{figure}

\paragraph*{S3 Fig.}
\label{sec:identifiability}
\textbf{Identifiability of parameters over time in the Mosquito inverse problem, the sine-shape temperature (\figurename~\ref{fig:identifiability}).}
This is achieved by expressing the system as a system of linear equations, the parameters as unknown variables, and analyzing the Reduced row-echelon form of the coefficient matrix, performed separately for each time $t$.
In the figure, identifiable parameters are defined  as free parameters which can get arbitrary values.
The values are rounded to 6-digit precision, aligning with the PINN's level of precision after training.
The figure explains the PINN's inaccuracies shown in \figurename~\ref{fig:inverse_t=sin}.

\begin{table}[H]
\begin{adjustwidth}{-2.00in}{0in}
\centering
\caption{ODE Model Parameters}
\label{tab:mosquito_ode_parameters}
\begin{tabular}{|c|l|c|l|}
\hline
Parameter      & Description                                 & Value & Unit          \\ \hline \hline
$\tau$            & Temperature                                 & & $^\circ C$ \\ \hline 
$\gamma_{Aem}$ & Development rate of emerging adults         & 1.143                                                                         & $days^{-1}$   \\ \hline
$\gamma_{Ab}$  & Development rate of bloodseeking adults     & 0.885                                                                         & $days^{-1}$   \\ \hline
$\gamma_{Ao}$  & Ovipositing adult development rate          & 2                                                                             & $days^{-1}$   \\ \hline
$f_E(>0)$      & Egg development rate                  & $0.16 \cdot \left(e^{\left[ 0.105(\tau-10) \right]} - e^{\left[ 0.105 (38 - 10) - \frac{1}{5.007}(38 - \tau) \right]} \right)$  & $days^{-1}$   \\ \hline
$f_P$          & Pupa development rate                 & $0.021 \cdot \left(e ^ {\left[ {0.162(\tau-10)} \right]} - e^{\left[ {0.162 (38 - 10) - \frac{1}{5.007}(38 - \tau) }\right])} \right)$ & $days^{-1}$   \\ \hline
$f_L$          & Larva development rate               & $ f_P / 1.65  $                                                            & $days^{-1}$   \\ \hline
$f_{Ag}(>0)$   & Development rate of gestating adults  & $\frac{\tau - 9.8}{64.4}$                         & $days^{-1}$   \\ \hline
$m_E$          & Egg mortality rate                      & $m_E = \mu_E$                                                                 & $days^{-1}$   \\ \hline
$m_L$          & Larval mortality rate                   & $ \exp \left[{-\tau/2 }\right]+\mu_L$                                                              & $days^{-1}$   \\ \hline
$m_P$          & Pupa mortality rate                     & $\exp \left[{-\tau/2 }\right]+\mu_P$                                                              & $days^{-1}$   \\ \hline
$m_A(>\mu_A)$  & Mortality rate of $Ab$,                 & $-0.005941 + 0.002965 \cdot \tau$                                                & $days^{-1}$   \\ \hline
$\mu_E$        & Minimum egg mortality rate                  & $0$                                                                           & $days^{-1}$   \\ \hline
$\mu_L$        & Minimum larval mortality rate               & $0.0304$                                                                      & $days^{-1}$   \\ \hline
$\mu_P$        & Minimum pupa mortality rate                 & $0.0146$                                                                      & $days^{-1}$   \\ \hline
$\mu_{em}$     & Mortality rate during emergence             & $0.1$                                                                         & $days^{-1}$   \\ \hline
$\mu_r$        & Mortality rate during bloodseeking          & $0.08$                                                                        & $days^{-1}$   \\ \hline
$\mu_A$        & Minimum adult mortality rate                & $\frac{1}{43}$                                                                & $days^{-1}$   \\ \hline
$\kappa_L$     & Carrying capacity for larvae                & $8 \cdot 10 ^ 8$                                                              & $days^{-1}$   \\ \hline
$\kappa_P$     & Carrying capacity for pupae                 & $10^7$                                                                        & $days^{-1}$   \\ \hline
$\sigma$       & Sex ratio at emergence                      & 0.5                                                                           & -             \\ \hline
$\beta$        & Number of eggs per $Ao$                     & $\beta_1=141 (np),  \beta_2 = 80 (p)$                                         & -             \\ \hline
\end{tabular}

\end{adjustwidth}
\end{table}

 \newcommand{\noop}[1]{}

%
%
%

\end{document}